\documentclass[aps,prb,reprint,showpacs,longbibliography,superscriptaddress]{revtex4-1}
\usepackage{amsmath,amssymb,bm}
\usepackage{graphicx}
\usepackage{latexsym}
\usepackage{color}
\usepackage{hyperref}

\newcommand{\DTN}{{NiCl$_{2}\cdot$4SC(NH$_2$)$_2$}}
\newcommand{\DTNX}{{NiCl$_{2-2x}$Br$_{2x}\cdot$4SC(NH$_2$)$_2$}}

\newcommand{\hamilt}{{{\mathcal{H}}}}

\newcommand{\aver}[1]{\left\langle #1 \right\rangle}

\begin{document}
\title{Dynamics of a bond-disordered $S=1$ quantum magnet near $z=1$ criticality}

\author{K.~Yu.~Povarov}
    \email{povarovk@phys.ethz.ch}
   \affiliation{Neutron Scattering and Magnetism, Laboratory for Solid State Physics, ETH Z\"{u}rich, Switzerland}
   \homepage{http://www.neutron.ethz.ch/}

\author{E.~Wulf}
    \affiliation{Neutron Scattering and Magnetism, Laboratory for Solid State Physics, ETH Z\"{u}rich, Switzerland}

\author{D.~H\"{u}vonen}
    \affiliation{National Institute of Chemical Physics and Biophysics, Akadeemia tee 23, 12618 Tallinn, Estonia}

\author{J.~Ollivier}
    \affiliation{Institut Laue-Langevin, 6 rue Jules Horowitz, 38042 Grenoble, France}

\author{A.~Paduan-Filho}
    \affiliation{High Magnetic Field Laboratory, University of S\~{a}o Paulo, 05315-970 S\~{a}o Paulo, Brazil}

\author{A.~Zheludev}
 \email{zhelud@ethz.ch}
 \affiliation{Neutron Scattering and Magnetism, Laboratory for Solid State Physics, ETH Z\"{u}rich, Switzerland}

\date{\today}

\begin{abstract}
Neutron scattering is used to study \DTNX, $x=0.06$, a
bond-disordered modification of the well-known gapped $S=1$
antiferromagnetic quantum spin system \DTN. The magnetic
excitation spectrum throughout the Brillouin zone is mapped out at $T=60$~mK using high-resolution time-of-flight spectroscopy.  It is found that the dispersion of spin excitation is renormalized, as compared to that in the parent compound. The lifetime of excitations near the bottom of the band is substantially decreased. No localized states are found below the gap energy $\Delta\simeq0.2$~meV. At the same time, localized zero wave vector states are detected above the top of the band. The results are consistent with a more or less continuous random distribution of bond strengths, and a discrete, possibly bimodal, distribution of single-ion anisotropies in the disordered material.
\end{abstract}



\maketitle

\section{Introduction}

Quantum magnets with disorder and randomness are presently
attracting a great deal of
attention.\cite{GlarumGeschwind_PRL_1991_HaldaneDoped,Fisher_PRB_1994_RandomSinglet,Vojta_JPA_2006_DisorderReview,ZheludevRoscilde_CRPhysique_2013_ReviewDirtyBosons}
They often exhibit behavior that is qualitatively different from
that of their disorder-free counterparts. There are actually several
ways to introduce randomness in a magnetic material. The effect of
\textit{site disorder} (or site dilution) is rather well understood.
Randomly removing spins has particularly severe consequences for
gapped quantum paramagnets. Their non-magnetic ground state is
destroyed, and magnetic order often emerges at finite temperatures
through the formation of correlated clusters around the
impurities.\cite{ShenderKivelson_PRL_1991_OrderSiteDisorder,Uchiyama_PRL_1999_PNVOorder,Uchinokura_JPCM_2002_CuGeO3review,SmirnovGlazkov_JETP_2007_disorderreview}
In contrast, the effect of random exchange interactions, or
\textit{bond disorder}, is more subtle. In systems with a spin gap
the ground state may survive. However, materials close to a magnetic
quantum-critical point (QCP) are more drastically affected. Indeed,
near QCP even a small variation in the Hamiltonian parameters may
lead to the qualitative change of the ground
state.\cite{Sachdev_2011_QPTBook,Sachdev_Science_2000_QCPs,Sachdev_NPhys_2008_QCPs}
Recently, Vojta\cite{Vojta_PRL_2013_InGap} has numerically
investigated the effect of bond disorder on the dynamic properties
of a gapped quantum magnet in the vicinity of the $z=1$
QCP.\footnote{{This $z=1$ QCP is not to be confused with the $z=2$
field-induced transition alike Bose--Einstein condensation of
magnons.}} Among the predictions are some intriguing
disorder-induced features, such as  localized states inside the gap,
or the possibility of ``weak ordering''.\footnote{This is a
coexistence of sharp Bragg peaks with extremely broadened
excitations at finite energies.} Unfortunately, despite the rapid
increase in the number of organometallic quantum
magnets,\cite{LandeeTurnbull_EuJInChem_2013_ReviewMagnets} and the
ease of achieving bond disorder via chemical substitution on
nonmagnetic sites involved in
superexchange,\cite{Yankova_PhilMag_2012_ReviewXtals} there is
presently a shortage of suitable model compounds to test the
predictions of Ref.~\onlinecite{Vojta_PRL_2013_InGap}. Most of the
known gapped quantum magnets (e.~g.
TlCuCl$_3$,\cite{OosawaTanaka_PRB_2002_TlCuCl3disorder}
IPA-CuCl$_{3}$,\cite{NafradiKeller_PRB_2013_IPAXblueshift}
PHCC,\cite{Huvonen_PRB_2012_PHCXdiffraction,Huvonen_PRB_2012_PHCXneutron,Huvonen_PRB_2013_PHCXphasediagram,Glazkov_JPCM_2014_PHCXesr}
Sul-Cu$_2$Cl$_4$,\cite{WulfMuhlbauer_PRB_2011_SulDisordered}
Cu(Qnx)Cl$_2$\cite{Keith_Polyhedron_2011_CQX,
PovarovLorenz_JMMM_2014_CQX}) exhibit a ``blue shift'' of magnons in
the presence of bond disorder. Thus, chemical modification increases
the spin gap and pushes these systems {\it away} from the $z=1$ QCP.

\begin{figure}[b]
  \includegraphics[width=0.5\textwidth]{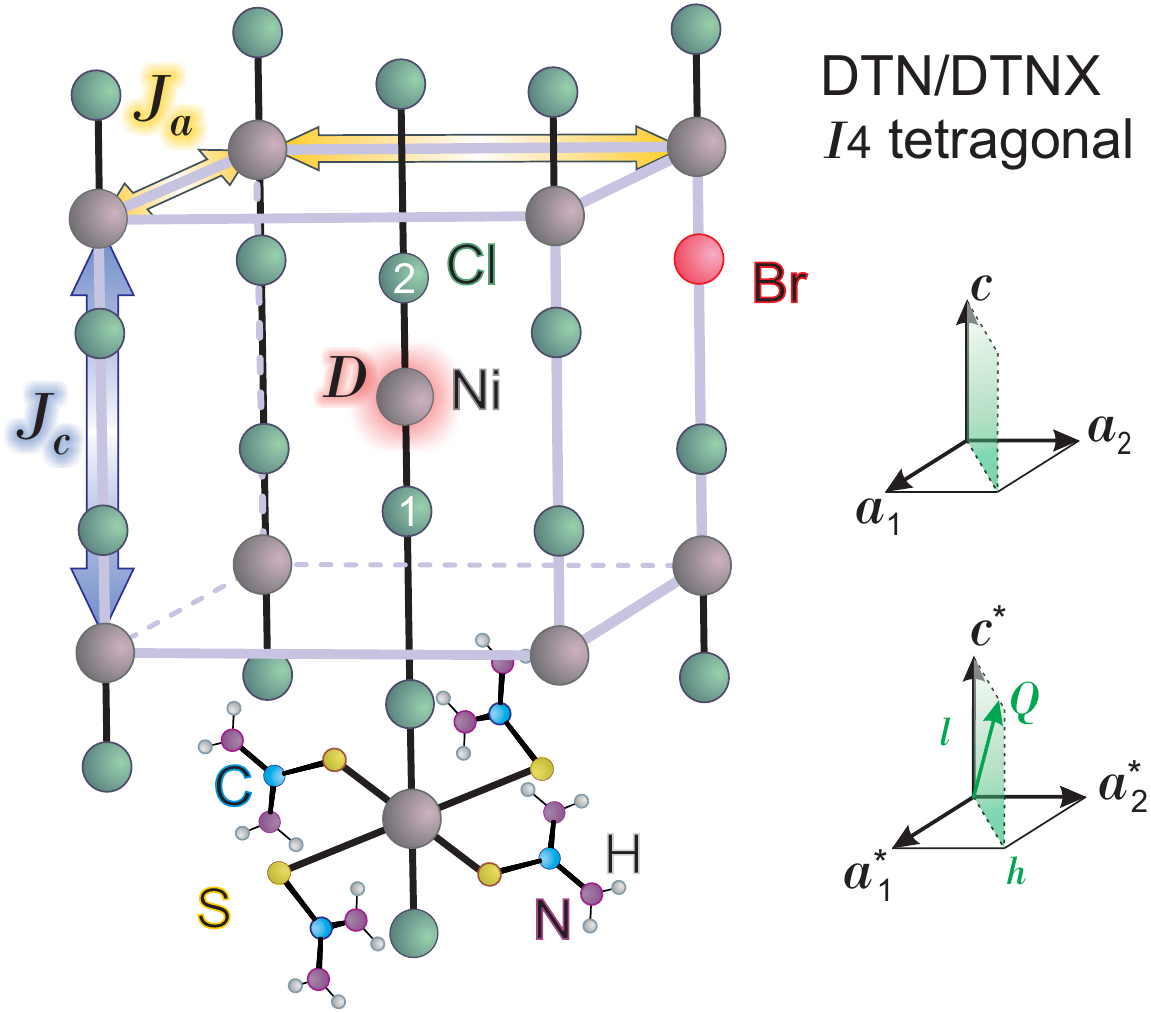}\\
  \caption{(Color online) A sketch of the \DTNX\ structure. The surrounding thiourea units are shown only for one of the Ni$^{2+}$ ions. The main magnetic
  interactions (see Hamiltonian (\ref{EQ:Hamiltonian})) for $S=1$ nickel ions are labeled. Two non-equivalent chlorine positions are marked as 1 and 2. The scattering plane for the present neutron experiment is shown on the right, in both real (top) and reciprocal (bottom) space representations.}\label{FIG:structure}
\end{figure}

One known exception is the extensively studied $S=1$ gapped quantum
magnet dichlorotetrakis-thiourea nickel \DTN\ (commonly abbreviated
as DTN).\cite{PaduanFilho_JChemPhys_1981_DTNfirst,
Zapf_PRL_2006_BECinDTN, Zvyagin_PRL_2007_ESRinDTN,
YinXia_PRL_2008_DTNcritical,Mukhopadhyay_PRL_2012_NMRinDTNandDIMPY,
Tsyrulin_JPCM_2013_DTNneutrons,
WulfHuvonen_PRB_2015_DTNIntrinsicBroadening} In contrast to
 integer-spin Heisenberg magnets such as
CsNiCl$_3$,\cite{Kenzelmann_PRL_2001_CsNiCl3,Zaliznyak_PRL_2001_CsNiCl3}
the gap in DTN is due to huge planar single-ion anisotropy. The
three-dimensional interactions in DTN are also strong enough to make
the Haldane gap physics
irrelevant.\cite{Haldane_PLA_1981_Gap,SakaiTakahashi_PRB_1990_DTNlikeGS}
Bond disorder is introduced into DTN by a random Br substitution on
the Cl site. This actually produces a ``red shift'' (decrease) of
the spin
gap.\cite{YuYin_Nat_2012_DTNboseglass,WulfHuvonen_PRB_2013_DTNXdiffraction}
In this respect, Br-substituted DTN, \DTNX\ (below abbreviated as
DTNX), appears to be the most promising candidate for an
experimental realization of the physics discussed in
Ref.~\onlinecite{Vojta_PRL_2013_InGap}.

The present work is a study of {spin dynamics} in DTNX with $x=6$\%
Br substitution. High-resolution inelastic neutron scattering
experiments provide a detailed picture of magnetic excitations
throughout the Brillouin zone. The measured spectra exhibit certain
key differences compared to the parent material: an increase of the
bandwidth, a reduction of excitation lifetimes, and new localized
states at high energies. We discuss the relation between these
experimental results and the numerical predictions of
Ref.~\onlinecite{Vojta_PRL_2013_InGap}. In addition, we touch on the
relevance of our findings to the previous thermodynamic studies of
DTNX in applied magnetic
fields.\cite{YuYin_Nat_2012_DTNboseglass,WulfHuvonen_PRB_2013_DTNXdiffraction}

The paper is organized as follows: In Sec.~\ref{SEC:experiment} we
review the structure and magnetic properties of our target material,
and describe the experimental setup and techniques; a brief overview
of the collected data is given in Sec.~\ref{SEC:data}, followed by
detailed analysis and discussion in Sec.~\ref{SEC:analysis};
Sec.~\ref{SEC:conclusive} summarizes our conclusions and outlines
the future research directions; some more technical issues are laid
out in the appendixes.

\section{Material and experimental details}
\label{SEC:experiment}

\subsection{The parent and disordered compounds}

Disorder-free DTN \DTN\ is an organometallic magnet belonging to a highly
symmetric $I4$ tetragonal space group.\cite{LopezCastroTruter_JChemS_1963_DTNxray} Magnetic $S=1$ Ni$^{2+}$ ions
occupy the positions in the body-centered tetragonal lattice (see
Fig.~\ref{FIG:structure}) with $a=9.558$~{\AA} and
$c=8.981$~{\AA}. Due to
this body-centered arrangement, the magnetic system can be seen as a
combination of two interpenetrating tetragonal
lattices that are totally decoupled at the mean-field level. Within each of these subsystems the magnetic properties
are captured by the following Hamiltonian:

\begin{equation}\label{EQ:Hamiltonian}
    \hamilt=\sum\limits_{\aver{i,j},n}D(S^{z}_{in})^{2}+J_{c}\textbf{S}_{in}\textbf{S}_{i(n+1)}+J_{a}\textbf{S}_{in}\textbf{S}_{jn},
\end{equation}

where $\aver{i,j}$ is the summation over the nearest neighbors in
the tetragonal $(a,a)$ plane, and index $n$ refers to the summation
along the $c$ axis. The magnetic properties are dominated by the
strong single-ion anisotropy of easy-plane type $D\simeq0.78$~meV.
The next term in the magnetic hierarchy is the nearest-neighbor
Heisenberg exchange interaction $J_{c}\simeq0.141$~meV within the
chains running along the high-symmetry axis. The least important
interaction is the interchain coupling $J_{a}\simeq0.014$~meV, which
is an order of magnitude weaker than
$J_{c}$.\cite{Zapf_PRL_2006_BECinDTN} As a result of the huge
anisotropy, the ground state of DTN is a quantum-disordered XY
paramagnet with a spin-singlet nonmagnetic ground state and a spin
gap $\Delta\simeq0.3$~meV.

It has been previously demonstrated that Br substitution has a very
slight effect on the lattice parameters and does not affect the
symmetry.\cite{YuYin_Nat_2012_DTNboseglass} The bromine ions are
site selective: They occupy Cl-1 positions in the lattice as shown
in Fig.~\ref{FIG:structure}. Chemical substitution decreases the
critical field $H_{c1}$ of magnetic ordering in DTNX,  indicating a
reduced spin
gap.\cite{YuYin_Nat_2012_DTNboseglass,WulfHuvonen_PRB_2013_DTNXdiffraction}
This gap reduction is significant: Already at $x=13$\% bromine
concentration $\Delta\simeq0.1$~meV is almost three times smaller
than in the parent compound.

\subsection{Quantities measured}
Magnetic inelastic neutron scattering almost directly probes the {\it dynamic spin structure factor}:

\begin{equation}
\label{EQ:DCF}
    \mathcal{S}^{\alpha\beta}(\mathbf{Q},\omega)=\int e^{-i[(\mathbf{Q}\cdot \mathbf{r})-\omega
    t]}\aver{S^{\alpha}(0,0)S^{\beta}(\mathbf{r},t)}\dfrac{d^{3}\mathbf{r}dt}{2\pi }.
\end{equation}
The actual neutron intensity measured in experiments is proportional to the differential cross section, and for unpolarized neutrons is given by: \cite{Squires_2012_Neutronbook}

\begin{equation}
    I(\mathbf{Q},\omega)\propto\frac{d^{2}\sigma}{dE d\Omega}\propto F^{2}(Q)\sum_{\alpha}\left(1-\frac{Q_{\alpha}^{2}}{Q^2}\right)\mathcal{S}^{\alpha\alpha}(\mathbf{Q},\omega).
\label{EQ:scatteringlaw}
\end{equation}

Here $\mathbf{Q}$ and $\hbar\omega$ the momentum and energy
transfers,  respectively. $F^{2}(Q)$ is the  magnetic form factor
for Ni$^{2+}$, known from numerical
calculations.\cite{Prince_2004_XtalTables} No absolute normalization
of the data was performed in this study.

\subsection{Experimental setup}

In the present work we used large (mass $\sim1$~g) fully deuterated single crystals of DTNX with 6\% Br substitution. They were grown from the solution by the same method as in previous works.\cite{PaduanFilho_PRB_2004_DTNinfield,
Zapf_PRL_2006_BECinDTN, YuYin_Nat_2012_DTNboseglass,
WulfHuvonen_PRB_2013_DTNXdiffraction}
The measurements were carried out on the high resolution cold neutron time-of-flight
(TOF) spectrometer IN5 at Institut
Laue--Langevin.\cite{OllivierMutka_JPSJ_2011_IN5} Sample environment was a $^3$He-$^4$He
dilution cryostat.
All data were collected on two co-aligned single crystals at $T=60$~mK,
which is much lower than all the relevant energy scales of DTN. The principal experimental scattering plane was $(1,-1,0)$, providing access to scattering vectors of type $(h,h,l)$  (see Fig.~\ref{FIG:structure}). Two data sets were collected, using incident neutron energies
$E_{\text{i}}=2.26$ and $1.17$~meV, respectively.
A Gaussian fit to the elastic incoherent scattering provided an estimate of the energy resolution: $\sqrt{8\ln 2}\sigma\simeq38~\mu$eV and
$16~\mu$eV full width at half height, respectively. In the high-resolution setup, incoherent elastic scattering has virtually no effect on the data collected above $\hbar \omega=25~\mu$eV energy transfer.
For each incident energy, the sample was rotated step-wise through the data collection, to fully
cover the first Brillouin zone.

\begin{figure*}[!t]
  \includegraphics[width=\textwidth]{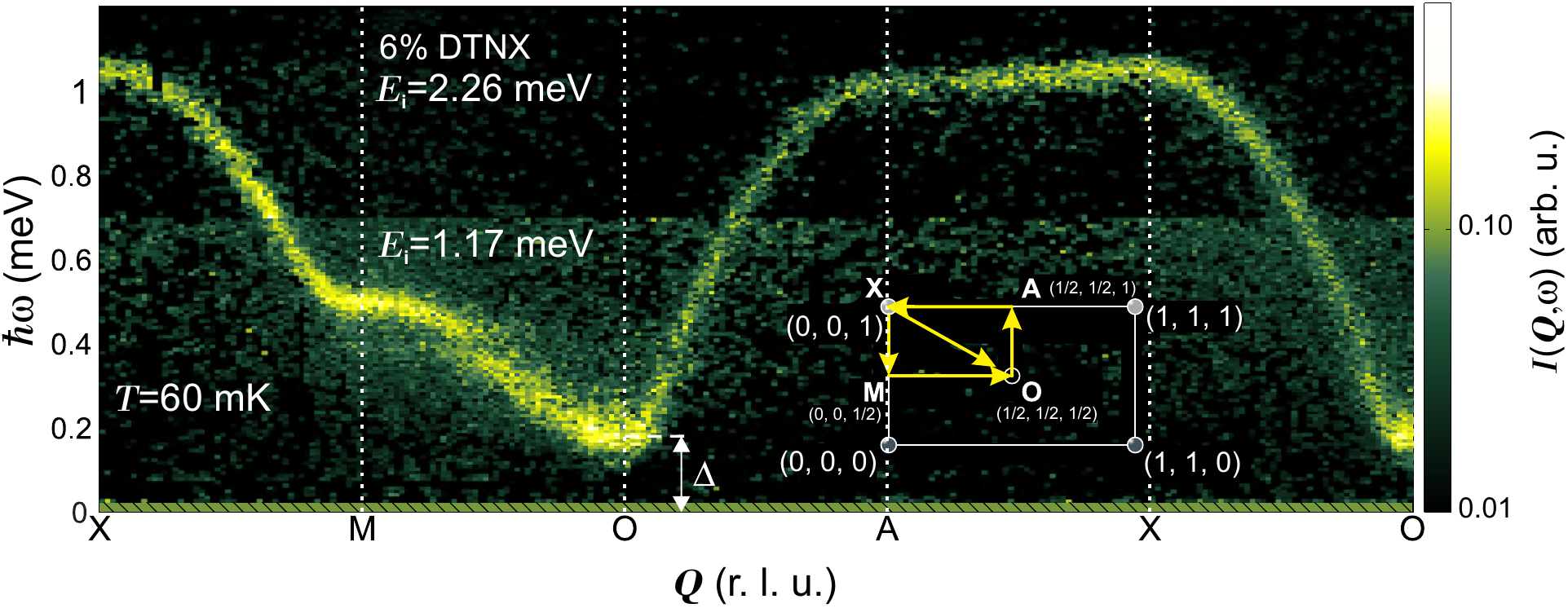}\\
  \caption{(Color online) False color plot of the neutron scattering intensity measured in 6\% DTNX at $T=60$~mK. The data are presented as a multi-panel cut along the trajectory shown in the inset. The data collected with $E_{\text{i}}=2.26$ and 1.17~meV incident energy are combined in each panel. The integration range in $\mathbf{Q}$ transverse to the cut directions is $\pm 0.02$ reciprocal lattice units.}\label{FIG:TetraZone_combo}
\end{figure*}

\section{Data overview}
\label{SEC:data}

The bulk of the neutron TOF data collected in 6\% DTNX at $T=60$~mK
is visualized in Fig.~\ref{FIG:TetraZone_combo}. These are a series
of false color plots of the measured inelastic intensity, presented
as two-dimensional momentum-energy cuts along high-symmetry
directions. The data below and above $0.7$~meV energy transfer  were
taken with neutron incident energies $E_{\text{i}}=1.17$~meV and
$E_{\text{i}}=2.26$~meV, respectively. We stress that the data are
plotted ``as is'', without any background subtraction.\footnote{The
uniform background is slightly higher in $E_{\text{i}}=1.17$~meV
dataset, and this is the origin of the visual ``cut-off'' at $0.7$~meV in Fig.~\ref{FIG:TetraZone_combo}.}

The measured spectrum of 6\% DTNX  is
qualitatively similar to that of disorder-free
DTN.\cite{Zapf_PRL_2006_BECinDTN} It is dominated by a well-defined magnon mode
with a dispersion minimum at the antiferromagnetic zone
center $\text{\textbf{O}}=(1/2, 1/2, 1/2)$. Two
saddle points are located at $\text{\textbf{M}}=(0, 0, 1/2)$ and $\text{\textbf{A}}=(1/2, 1/2, 0)$, correspondingly. The highest energy of the single-magnon
excitation is found at $\text{\textbf{X}}=(0, 0, 0)$. The main difference with the parent material
is a much reduced spin gap $\Delta\simeq0.2$~meV, compared to $\Delta\simeq0.3$~meV in DTN.

At lower energies, the excitations in DTNX appear less sharp than at
high energy transfers. Notably, there are no additional excited
states below the gap energy  at the \textbf{O} point or any other
region of the Brillouin zone down to $25$~$\mu$eV. Finally, we note
a subtle drop of intensity around $0.4$~meV. The latter is found in
data sets with both neutron incident energies, and hence is unlikely
to be of instrumental origin. Although it resembles an avoided
crossing in the false-color plot shown, it is actually always
observed at the same energy transfer, irrespective of wave vector.
Its origin remains unclear.

\section{Analysis and discussion}
\label{SEC:analysis}
    \subsection{Dispersive excitations}
    \label{SEC:analysisSMA}

\begin{figure}
  \centering
  \includegraphics[width=0.45\textwidth]{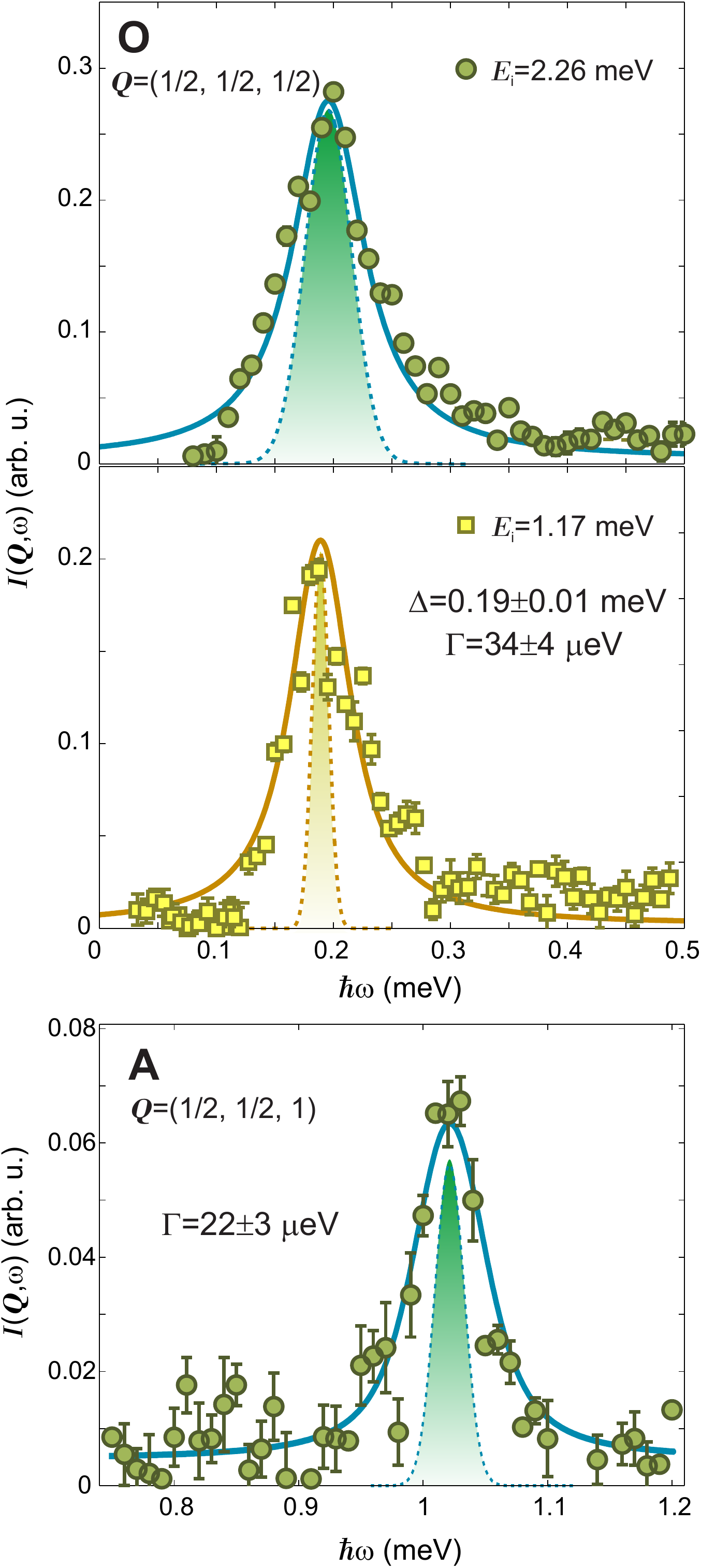}\\
  \caption{(Color online) Constant-$\mathbf{Q}$ cuts at the center (\textbf{O}) and at the boundary (\textbf{A})
  of antiferromagnetic Brillouin zone of 6\% DTNX. Squares and circles correspond to $E_{\text{i}}=1.17$~meV and $E_{\text{i}}=2.26$~meV,
  respectively. Voigt fits~(\ref{EQ:Voigt}) are shown by solid lines.
  The shaded areas represent the experimental energy resolution.}
  \label{FIG:1DcutsOA}
\end{figure}

\begin{table*}
  \centering
  \begin{tabular}{l c l c l c l c l c l}
  \hline\hline
   & & \multicolumn{1}{c}{DTN} & & \multicolumn{7}{c}{6\% DTNX}\\
   \cline{3-3}\cline{5-11}
    & & \begin{tabular}{@{}c@{}}Zapf \\ \emph{et al.} (2006)\cite{Zapf_PRL_2006_BECinDTN}\end{tabular} & & \begin{tabular}{@{}c@{}}Present study \\ Eq.~(\ref{EQ:Hamiltonian}) only\end{tabular} & & \begin{tabular}{@{}c@{}}Present study\\ Eq.~(\ref{EQ:Hamiltonian})$+J_{d}$\end{tabular} & & \begin{tabular}{@{}c@{}}Present study\\ Eq.~(\ref{EQ:Hamiltonian})$+J_{c2}$\end{tabular} & & \begin{tabular}{@{}c@{}}Present study\\ Eq.~(\ref{EQ:Hamiltonian})$+J_{d}+J_{c2}$\end{tabular}\\
   \hline
  $D$ & & $0.780(3)$ meV & & $0.792(3)$ meV & & $0.812(2)$ meV & & $0.805(2)$ meV & & $0.807(2)$ meV \\
  $J_{c}$ & & $0.141(3)$ meV & & $0.155(1)$ meV & & $0.158(1)$ meV & & $0.149(1)$ meV & & $0.150(1)$ meV \\
  $J_{a}$ & & $0.014(1)$ meV & & $0.0158(3)$ meV & & $0.0161(2)$ meV & & $0.0155(2)$ meV & & $0.0157(2)$ meV \\
  $J_{d}$ & & -& & - & & $0.0059(5)$ meV & & - & & $0.0060(4)$ meV \\
  $J_{c2}$ & & - & & - & & - & & $-0.0095(7)$ meV & & $-0.0096(7)$ meV \\
  \hline
    $D/J_{c}$ & & $5.5(1)$ & & $5.11(3)$ & & $5.14(3)$ & & $5.40(3)$ & & $5.38(3)$ \\
  $J_{a}/J_{c}$ & & $0.098(7)$ & & $0.102(2)$ & & $0.102(2)$ & & $0.104(2)$ & & $0.105(2)$ \\
  \hline\hline
\end{tabular}
  \caption{Results of fitting the RPA dispersion relation~(\ref{EQ:LinSW}) to the measured excitation energies in 6\% DTNX. Additional perturbations to the basic Hamiltonian~(\ref{EQ:Hamiltonian}) are considered (see Appendix~\ref{APP:perturbations} for more details). The results of an analogous analysis for disorder-free DTN from Ref.~\onlinecite{Zapf_PRL_2006_BECinDTN} are also shown.}
  \label{TAB:parameters}
\end{table*}

\begin{figure}
  \centering
  \includegraphics[width=0.5\textwidth]{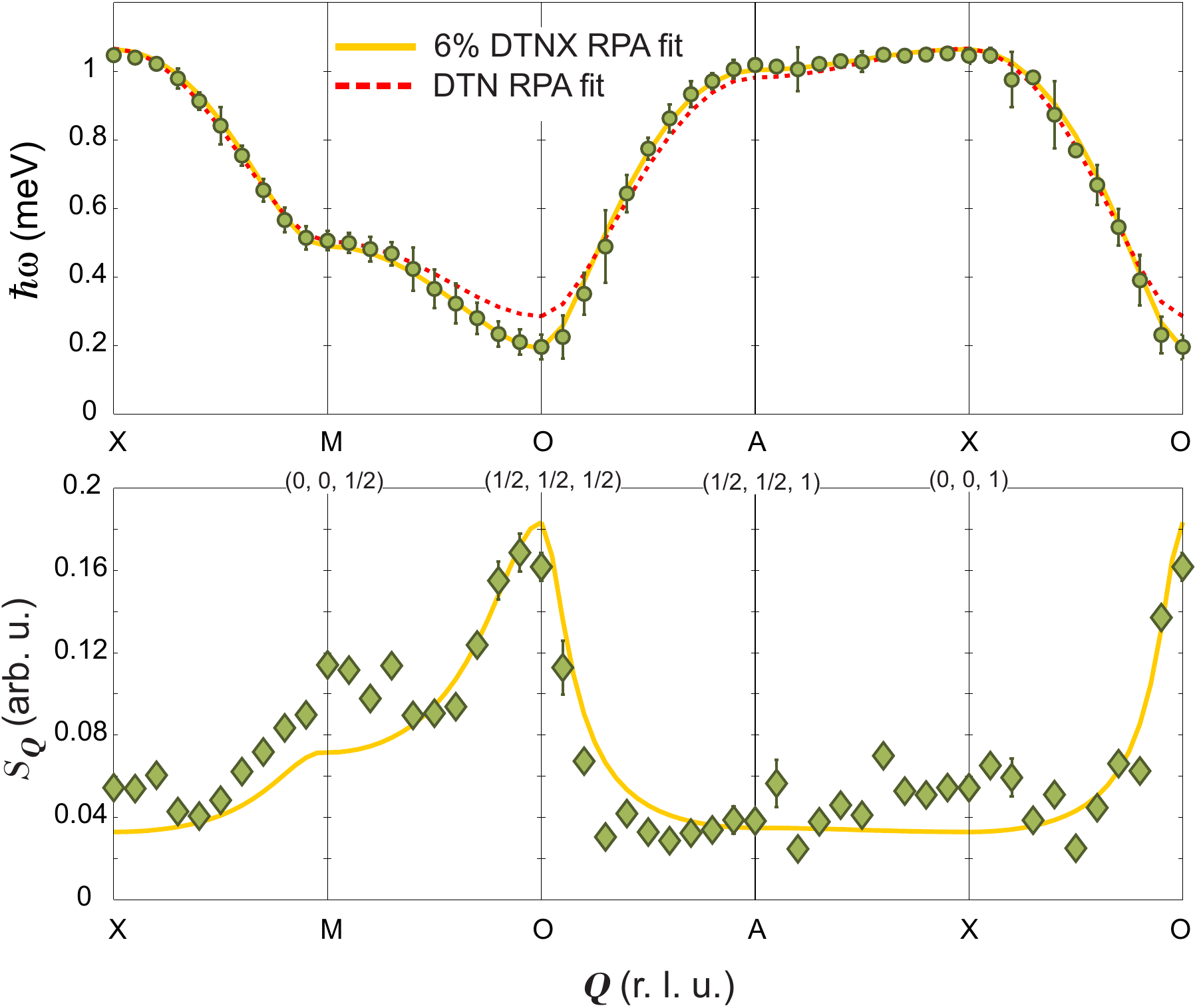}\\
  \caption{(Color online) Top panel: Magnon dispersion relation measured in 6\% DTNX, as deduced from Voigt fits to
  individual constant-$\mathbf{Q}$ cuts of $E_{\text{i}}=2.26$~meV dataset.
  The solid line is a fit to the RPA prediction of Eq.~(\ref{EQ:LinSW}) with the parameters summarized in the last column of Table~\ref{TAB:parameters}.
  The dashed line is the reference magnon dispersion in disorder-free DTN following Ref.~\onlinecite{Zapf_PRL_2006_BECinDTN}.
  Bottom panel: Integrated intensity corrected for the magnetic form factor and polarization factor, $\mathcal{S}_{\mathbf{Q}}$, as determined in fits
  to the same individual cuts. The solid line is the RPA result, which is simply the inverse of $\hbar\omega_{\mathbf{Q}}$.}
  \label{FIG:dispersion}
\end{figure}

Since we lack a suitable microscopic model to globally describe the
spectrum in a quantum magnet with disorder, we chose a more
empirical approach to data analysis. We quantitatively analyze the
$E_{\text{i}}=2.26$~meV dataset, covering the full magnetic
excitation band. The measured data were broken up into a series of
individual constant-$\mathbf{Q}$ cuts. In each such cut, the
scattering is a well-defined peak that we approximated by a Voigt
profile:

\begin{equation}\label{EQ:Voigt}
    I(\mathbf{Q},\omega)= \mathcal{S}_{\mathbf{Q}}F^{2}(Q)\left(2-\frac{Q_{\bot}^{2}}{Q^2}\right)V(\hbar\omega-\hbar\omega_{\mathbf{Q}},\sigma,\Gamma_{\mathbf{Q}}).
\end{equation}

The Gaussian width $\sigma$ represents experimental energy
resolution (the values quoted above, with the energy transfer
correction
included\cite{Lowde_JNE_1960_TOFresolution,Lechner_1985_TOFresolution}).
For each cut, the parameters of this model are as follows.
$\hbar\omega_{\mathbf{Q}}$ is the position of the peak that we
associate with the single-magnon energy. The Lorentian width
$\Gamma_{\mathbf{Q}}$ of the Voigt function represents the intrinsic
magnon line width and, potentially, wave vector resolution
(``focusing'') effects. $\mathcal{S}_{\mathbf{Q}}$ is the peak's
integrated intensity.  Due to the planar nature of the ground state,
we assumed that the observed magnon corresponds to transverse spin
fluctuations
$\mathcal{S}^{xx}(\mathbf{Q},\omega)=\mathcal{S}^{yy}(\mathbf{Q},\omega)$.
The polarization factors in the expression above are chosen
accordingly, with $Q_{\bot}$ being the momentum transfer in the
$(a,a)$ plane. Examples of fits to individual cuts can be found in
Fig.~\ref{FIG:1DcutsOA}.

The thus obtained magnon dispersion relation is plotted in symbols in the upper panel
of Fig.~\ref{FIG:dispersion}. These data were further analyzed within the
random-phase approximation
(RPA),\cite{JensenMackintosh_1991_RareEarthBook} treating the Heisenberg exchange as a perturbation to decoupled $S=1$ single ions. The RPA dispersion relation is:

\begin{equation}
 \label{EQ:LinSW}
    \hbar\omega_{\mathbf{Q}}=\sqrt{D^{2}+4D\gamma(\mathbf{Q})}.
\end{equation}

Here
$\gamma(\mathbf{Q})=\sum\limits_{\mathbf{r}}J_{\mathbf{r}}\cos(\mathbf{Qr})$
is essentially the Fourier transform Heisenberg exchange
interactions in the system. For our fit, we consider not only the
parameters $D$, $J_{c}$ and $J_{a}$, but also two possible
perturbations to the Hamiltonian~(\ref{EQ:Hamiltonian}): diagonal
exchange between the tetragonal sublattices $J_{d}$ and the
next-nearest neighbor exchange along the $c$ direction $J_{c2}$. The
reasons to introduce these perturbations are discussed in
Appendix~\ref{APP:perturbations}. Fitting the RPA dispersion
relation to the data in Fig.~\ref{FIG:dispersion} (top panel) yields
the parameter values summarized in Table~\ref{TAB:parameters}. The
fit itself is represented by the solid line. For a direct
comparison, we also quote Hamiltonian parameters for disorder-free
DTN from the previous studies\cite{Zapf_PRL_2006_BECinDTN} and plot
the corresponding dispersion relation in a dashed line.

As the analysis shows, the gap softening in DTNX is primarily due to
the increase of the bandwidth. Even as the anisotropy becomes
stronger with the Br substitution, the ratio $D/J_{c}$  actually
decreases from $5.5$ in pure DTN to $5.1-5.4$ in 6\% DTNX, depending
on the terms included in the model Hamiltonian. The ratio
$J_{a}/J_{c}$ is still around $0.1$ within the uncertainty limit.
The RPA dispersion~(\ref{EQ:LinSW}) can be modified to account for
quantum renormalization  of the bare Hamiltonian parameters. This
correction, described in Appendix~\ref{APP:RPArenorm}, changes the
$D/J_{c}$ ratio to $4.7$ in the parent
material.\cite{Zapf_PRL_2006_BECinDTN,ZhangWierschem_PRB_2013_DTNdispersion}
In 6\% DTNX, as found from the present experiment, it will be
between $4.06$ and $4.36$. The ratio $J_{a}/J_{c}\simeq0.1$ is
unchanged. Thus the reduced $D/J_{c}$ ratio is the key ingredient of
the excitations ``red shift'' in DTNX.

The intensities $\mathcal{S}_{\mathbf{Q}}$ obtained in fits to individual cuts are plotted in symbols in the lower panel of Fig.~\ref{FIG:dispersion}. In the RPA, $\mathcal{S}_{\mathbf{Q}}$
is simply proportional to $1/\hbar\omega_{\mathbf{Q}}$. As
shown by the solid line,  this expectation is in a reasonable agreement with the data. The remaining
discrepancies might be attributed to the possible contribution of
longitudinal fluctuations, not included in our approach.

\subsection{Approaching criticality}
\label{SEC:analysisCritical}

\begin{figure}
\centering
  \includegraphics[width=0.5\textwidth]{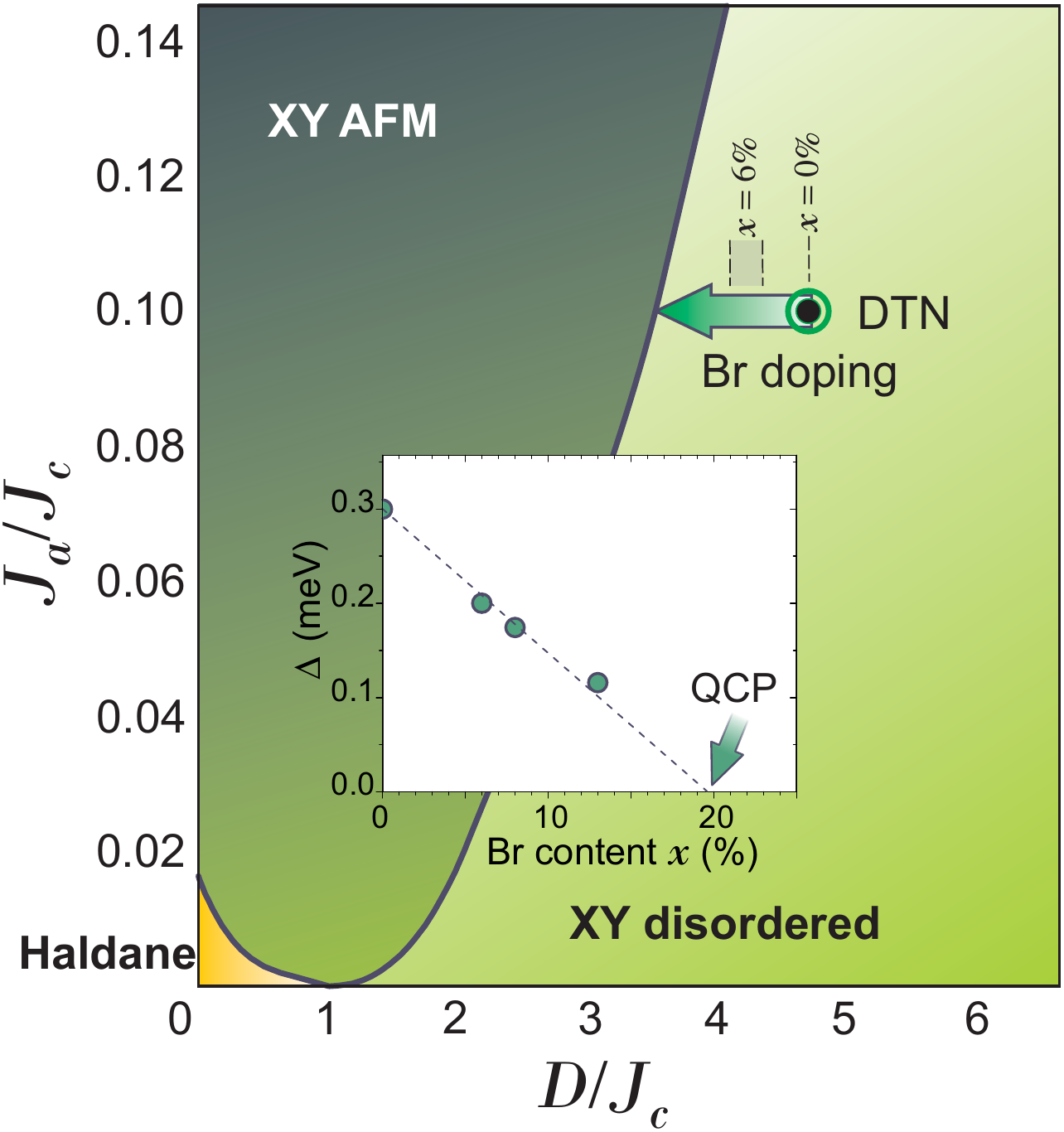}\\
  \caption{(Color online) Ground states of  Hamiltonian~(\ref{EQ:Hamiltonian}), based upon the results of Refs.~\onlinecite{WierschemSengupta_PRL_2014_DTNlikeGS,WierschemSengupta_MPhysLettB_2014_DTNlikeGS}. The arrow indicates the approximate trajectory of DTNX upon Br substitution. The inset shows the energy gap in DTNX as a function of bromine content. Data points are the result of the present  and previous studies.\cite{Zapf_PRL_2006_BECinDTN,YuYin_Nat_2012_DTNboseglass,WulfHuvonen_PRB_2013_DTNXdiffraction} The dashed line is a guide for the eye.}\label{FIG:qcp}
\end{figure}

Due to a random distribution of Br substitutes in the sample, the
parameters of the microscopic Hamiltonian, including exchange
constants and anisotropy, will themselves vary from one unit cell to
the next in a random manner. This said, it stands to reason that the
values obtained from analyzing magnon dispersion curves correspond
to the {\it average} values of these parameters. The main mechanism
leading to a decrease of the spin gap and driving DTNX closer to the
QCP is a then  a steadily decreasing $D/J_{c}$ ratio.

Fortunately, the phase diagram of Hamiltonian~(\ref{EQ:Hamiltonian})
is well known
numerically,\cite{SakaiTakahashi_PRB_1990_DTNlikeGS,WierschemSengupta_PRL_2014_DTNlikeGS,WierschemSengupta_MPhysLettB_2014_DTNlikeGS}
and shown in Fig.~\ref{FIG:qcp}. For small anisotropy and almost
isolated chains it includes a gapped topological Haldane phase.
Sufficiently strong interchain interactions restore XY-like long
range magnetic order. However, the system is again in a gapped
non-magnetic ``single ion'' state for large anisotropy. The values
of Hamiltonian parameters determined for the parent DTN
compound,\cite{Zapf_PRL_2006_BECinDTN} allow us to place it in the
latter region of the phase diagram (see Fig.~\ref{FIG:qcp}). As
mentioned above, Br substitution does not change the $J_a/J_c$ ratio
significantly. Instead, by decreasing $D/J_{c}$, the system is
driven left on the phase diagram, towards the line of long-range XY
ordering. The inset of Fig.~\ref{FIG:qcp} show the energy gap
$\Delta$ in DTNX as a function of Br content, as deduced from the
current and previous
studies.\cite{Zapf_PRL_2006_BECinDTN,YuYin_Nat_2012_DTNboseglass,WulfHuvonen_PRB_2013_DTNXdiffraction}
The decrease is roughly linear, and the gap may be expected to close
around 20\% Br content. At present it is not clear whether DTNX
samples with such high Br concentrations are stable.

\subsection{Magnon lifetimes}
\label{SEC:analysisWidth}

\begin{figure}
  \centering
  \includegraphics[width=0.5\textwidth]{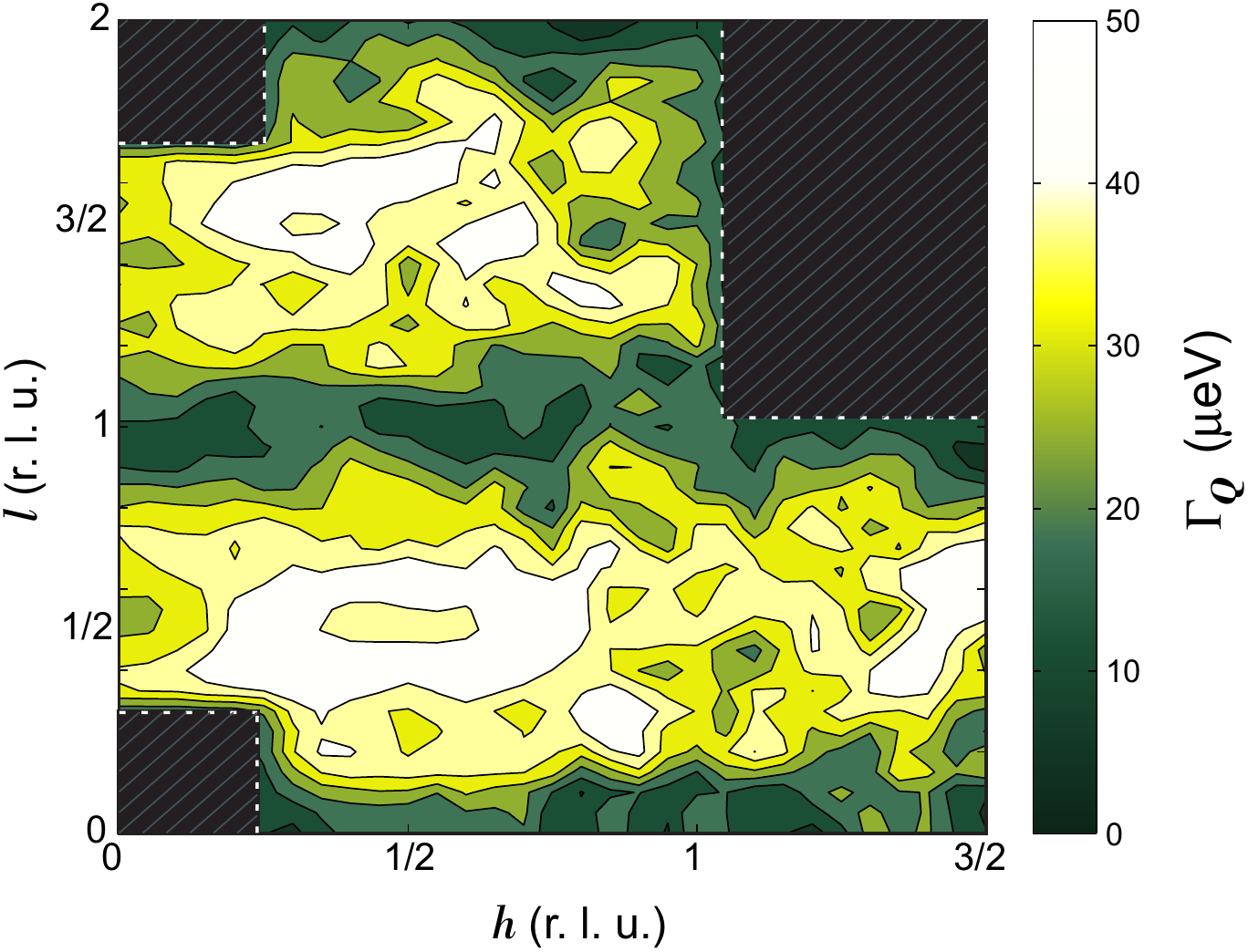}\\
  \caption{(Color online) False color map of measured intrinsic Lorentian magnon linewidth $\Gamma_{\mathbf{Q}}$ as found from Voigt fits to individual constant-$\mathbf{Q}$ cuts ($E_{\text{i}}=2.26$~meV dataset). Regions of reciprocal space with no data are shown hatched.}
  \label{FIG:linewidthRPA}
\end{figure}

Beyond a simple change of average Hamiltonian parameters, several
key features of the spectrum are to be attributed to disorder, {\it
i.e.}, to a microscopic random variation of these parameters in the
sample. As shown in Fig.~\ref{FIG:1DcutsOA}, the excitations at the
antiferromagnetic zone center \textbf{O} have a significant energy
width beyond the resolution of the instrument. This is further
emphasized by comparing data obtained with different neutron
incident energy. For $E_{\text{i}}=1.17$~meV the energy resolution
is much sharper than for $E_{\text{i}}=2.26$~meV, but the obtained
broad peak at $\mathbf{Q}=(1/2,1/2,1/2)$ is similar in both
cases, with effective
Lorentian linewidth $\Gamma=34\pm4$~$\mu$eV. Although at high
energies the excitations are sharper, their linewidth is still
pronounced. Fitted linewidth at, for instance, point \textbf{A} is
$\Gamma=22\pm3$~meV. As the resolution function of a time-of-flight
spectrometer sharpens with the increase in energy
transfer,\cite{Lowde_JNE_1960_TOFresolution,Lechner_1985_TOFresolution}
the linewidth $\Gamma$ still exceeds the estimated instrument
resolution $\sigma$ at $\hbar\omega\simeq1$~meV almost twice.

Wavevector resolution effects alone can not account for the observed
increase of linewidth. A conservative estimate of
$\mathbf{Q}$-resolution contribution to the peak width in energy
cuts can be done as
$\sigma_{\mathbf{Q}}\lesssim|\hbar\omega_{\mathbf{Q}+\delta
\mathbf{Q}}-\hbar\omega_{\mathbf{Q}}|$. In our case, for the
antiferromagnetic zone center this gives only
$\sigma_{\mathbf{Q}}\simeq5$~$\mu$eV, which is much less than the
observed broadening.

The contour map of measured intrinsic line width $\Gamma_{\mathbf{Q}}$ is
shown in Fig.~\ref{FIG:linewidthRPA}. One can see that
$\Gamma_{\mathbf{Q}}$ varies significantly along the
tetragonal axis, but has a less pronounced variation in a transverse
direction. The broadest excitations are found around the
antiferromagnetic zone center where the dispersion is a minimum. This is in a strong contrast with the picture of excitation
broadening previously observed in another bond-disordered gapped
antiferromagnet PHCX.\cite{Huvonen_PRB_2012_PHCXneutron} In the latter
case, magnon damping is due to scattering on isolated (discrete) impurities. As a result, it roughly scales with the single-magnon density of states, and is a maximum at the band top. In DTNX we find exactly the opposite. As will be argued below, the
observed behavior of $\Gamma_{\mathbf{Q}}$ is consistent with a
smooth continuous distribution of Heisenberg exchange strengths in the
material.

\subsection{Localized states}
\label{SEC:analysisSaucers}

\begin{figure}
\centering
  \includegraphics[width=0.5\textwidth]{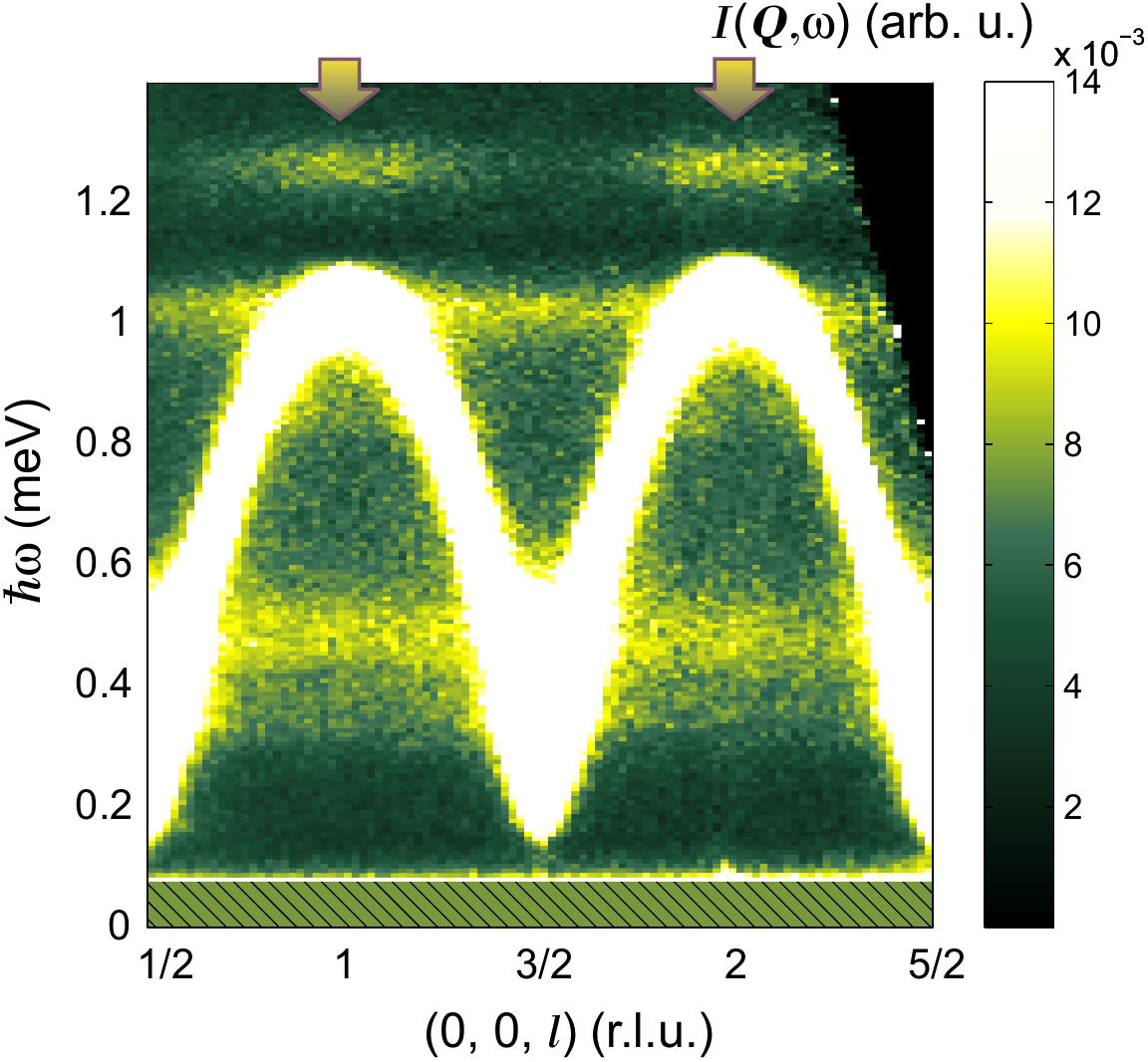}\\
  \caption{(Color online) High contrast false color map of $I(\mathbf{Q},\omega)$ for $\mathbf{Q}$ along $\mathbf{c}^{\ast}$ (here $E_{\text{i}}=2.26$~meV). The integration range in transverse direction is the entire Brillouin zone. Note the dispersionless (local) excitations just above the top of the magnon band denoted by arrows.}\label{FIG:saucers}
\end{figure}

Figure~\ref{FIG:saucers} shows a high-contrast false color plot of
$I(\mathbf{Q},\omega)$ integrated over the whole Brillouin zone and
projected onto a $\mathbf{c}^{\ast}$ direction. Note that here we
use a linear intensity scale, compared to the logarithmic scale used
above in Fig.~\ref{FIG:TetraZone_combo}. In Fig.~\ref{FIG:saucers},
the magnon band looks completely ``overexposed''. There are clearly
{\it no additional states visible inside the gap}. However,
additional states are found just above the top of the band at
$\hbar\omega\simeq1.25$~meV. These are {\it local} excitations, as
they show no dispersion.

Given the low intensity of the feature in 6\% DTNX, how reliable is
this observation? An important argument here is that the feature at
$1.25$~meV has an intensity distribution, perfectly matching the
sample's reciprocal lattice. The intensity is cosine modulated along
$[0~0~1]$ and constant along $[1~1~0]$. This is nontypical for the
spurious features usually having no
$\mathbf{Q}$-structure.\cite{Pintschovius_JAC_2014_18mevSpurion} As
the numerical calculations show, localized feature of this kind can
actually be expected in a bond-disordered
system\cite{Vojta_PRL_2013_InGap,UtesovSizanov_PRB_2014_DisorderExcitations}
(this will be discussed in more details below). Indirect evidence of
high-energy states in DTNX also comes from the experimentally
observed pre-saturation ``pseudoplateau'' in
magnetization.\cite{YuYin_Nat_2012_DTNboseglass} The width of the
plateau $\Delta H\simeq1.5$~T and the observed energy separation
between the band top and the localized state
$\hbar\Delta\omega\simeq0.2$~meV are in a rough agreement. As these
states are located at the regions of reciprocal space, equivalent to
$Q=0$ momentum transfer, they should also be observable by such
techniques as electron spin resonance and THz spectroscopy. A very
recent THz spectroscopy experiment does show the presence of an
additional feature at approximately the same energy in DTNX and its
absence in the parent material.\footnote{{D. H\"{u}vonen \emph{et
al.}, unpublished}}

\subsection{Disorder analysis}
\label{SEC:analysisBonds}

\begin{figure}
\centering
  \includegraphics[width=0.5\textwidth]{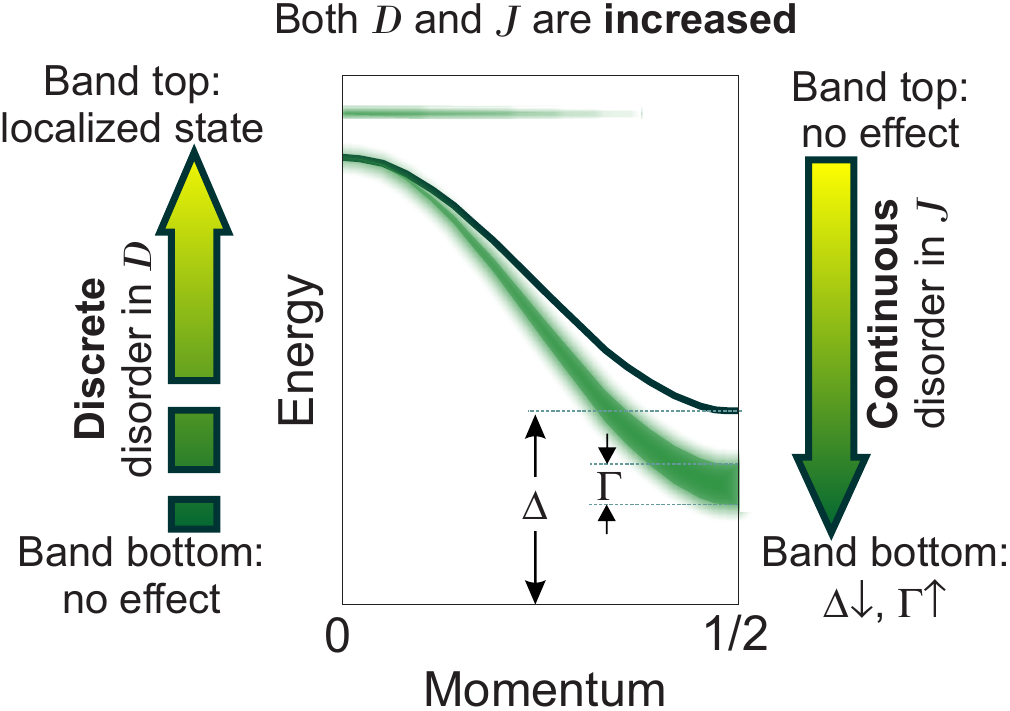}\\
  \caption{(Color online) A cartoon highlighting the main effects of various kinds of disorder on the spin excitation spectrum. The solid line is the magnon dispersion  in the absence of disorder. The blurred curve is the spectrum of the disordered system.}\label{FIG:memo}
\end{figure}

To understand the emergence of local excitations and other features
of the observed spectrum, we can establish a crude qualitative
``mapping'' between the Hamiltonians of DTN~(\ref{EQ:Hamiltonian})
and that of the dimer modeled studied in
Ref.~\onlinecite{Vojta_PRL_2013_InGap}. The latter has two
parameters: the intradimer exchange constant $\mathcal{J}$ and the
interdimer coupling $\mathcal{K}$. The dimer strength $\mathcal{J}$
primarily determines the gap and interdimer exchange $\mathcal{K}$
determines the bandwidth. The increase of the former drives the
system away, and of the latter --- closer to QCP.  In this sense,
our $S=1$ individual magnetic sites with single-ion anisotropy can
be seen as spin-gap objects analogous to the $S=1/2$ dimers. The
parameter $\mathcal{J}$ is then naturally mapped to the single-ion
anisotropy $D$ and the critical coupling $\mathcal{K}$ corresponds
to the Heisenberg exchange $J$. Similar mapping between dimerized
and single-ion anisotropy systems has also been used in a recent
theoretical study of Utesov \textit{et
al}.\cite{UtesovSizanov_PRB_2014_DisorderExcitations}

The numerical study Ref.~\onlinecite{Vojta_PRL_2013_InGap} predicts
the presence of in-gap states in two cases of \textit{discrete}
disorder distribution: a small fraction of isolated sites having
$\mathcal{J}<\aver{\mathcal{J}}$  (in our correspondence with DTN
$D<\aver{D}$)\footnote{Ref.~\onlinecite{Vojta_PRL_2013_InGap},
Fig.~1} or a small fraction of sites with
$\mathcal{K}>\aver{\mathcal{K}}$ (in our case
$J>\aver{J}$).\footnote{Ref.~\onlinecite{Vojta_PRL_2013_InGap}
Supplementary Material, Fig.~S2} Our analysis of the magnon spectrum
in DTNX shows that anisotropy $D$ is effectively increased by Br
substitution, so the former scenario is clearly not applicable.
Thus, not having any in-gap bound states in DTNX may indicate that
the disorder of the exchange constant $J$ is either weak or
non-discrete. At the same time, as captured by the simulations of
Ref.~\onlinecite{Vojta_PRL_2013_InGap} (Fig.~5 therein), a small
number of sites with a singular large value of
$\mathcal{J}>\aver{\mathcal{J}}$ (i. e. $D>\aver{D}$) produces a
localized state near the top of the magnon band. This is totally in
line with our observation of high-energy local excitations in DTNX.
The underlying assumption is that the anisotropy distribution is
{\it discrete}, with just a few Ni$^{2+}$ ions having a
substantially increased $D$-term. The simulations also predict a
broadening of the main magnon branch, in the case of a {\it
continuous} broad distributions of either $\mathcal{K}$ and
$\mathcal{J}$. The former broadens the magnons evenly in the entire
Brillouin zone (Ref.~\onlinecite{Vojta_PRL_2013_InGap} Supplementary
Material, Fig.~S4), while the latter mostly affects magnons at the
bottom of the band (Fig.~S5). Comparing it to our observations, we
may again guess that the exchange constants in DTNX show a rather
broad continuous distribution.

Certainly, the above discussion is based on a rather tenuous analogy between two very different Hamiltonians. However, if we accept this qualitative correspondence, the following picture emerges. The {\it average} values of both $\langle D \rangle $ and $\langle J\rangle$ are increased
with the Br substitution, modifying the magnon dispersion relation accordingly. The anisotropy is decreased on only a few sites, probably in direct proximity of the Br substitutes. The exchange constants, however, have a broad statistical distribution around the average value. Schematically, this scenario is borne out in Fig.~\ref{FIG:memo}. A broad distribution of exchange constants in DTNX is not unexpected. It results from each Br substitute affecting a large number of bonds. Each such size-mismatched Br defect creates a strain field in the crystal, which, in turn, affects bond angles and thereby superexchange interactions.  The strain field falls off as  $\sim r^{-3}$ (Ref.~\onlinecite{IndenbomLothe_1992_DefectBook}), and in a soft material like DTNX is expected to be rather long-range.

\section{Conclusions}
\label{SEC:conclusive}

In summary,  Br substitution has a profound effect on the spin dynamics of DTN, even at rather low concentrations. On the one hand,  both anisotropy and exchange interactions are, on the average, increased. The ratio $D/J_{c}$ actually decreases, reducing the spin gap and driving the system closer to the QCP. Simultaneously, new features emerge due to disorder. Magnon lifetimes are shortened, predominantly at the bottom of the band. Somewhat counter-intuitively, localized states appear not inside the spin gap, but just above the top of the magnon band. This behavior can be explained by a hand-waving analogy with disorder in the Heisenberg-dimer model.\cite{Vojta_PRL_2013_InGap}
We hope that our experiments will stimulate a study of randomness in the anisotropic single-ion Hamiltonian appropriate for DTN.

\acknowledgements This work was supported by the Swiss National
Science Foundation, Division~2, the Estonian Ministry of Education
and Research under Grant No. IUT23-03 and the Estonian Research
Council Grant No. PUT451. AP-F acknowledges support from
CNPq-Brazil. We would like to thank Dr. D. Schmidiger for helpful
discussions and Dr. S. Gvasaliya for assistance with the sample
preparation.

\appendix

\section{Additional interactions}\label{APP:perturbations}

\begin{figure}
\centering
  \includegraphics[width=0.5\textwidth]{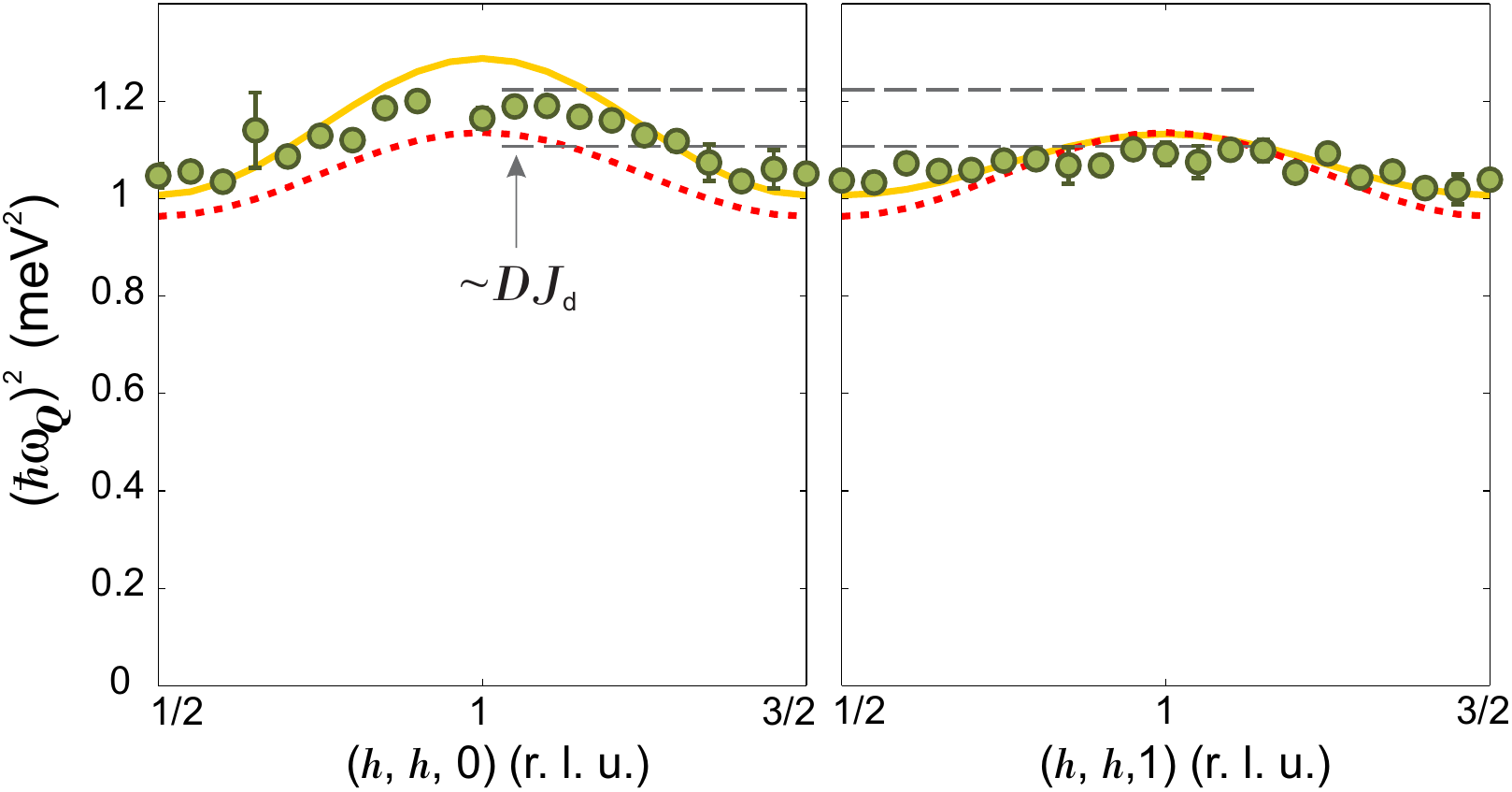}\\
  \caption{(Color online) Symbols: squared dispersion $(\hbar\omega_{\mathbf{Q}})^{2}$ along the $(h,h,0)$ and $(h,h,1)$ directions ($E_{\text{i}}=2.26$~meV). The dashed line is the dispersion model for pure DTN used by Zapf \emph{et al.}\cite{Zapf_PRL_2006_BECinDTN}. The solid line is best fit with $J_{d}$ perturbation included, as in the last column of Table~\ref{TAB:parameters}.}\label{FIG:hw2}
\end{figure}

\begin{figure}
\centering
  \includegraphics[width=0.5\textwidth]{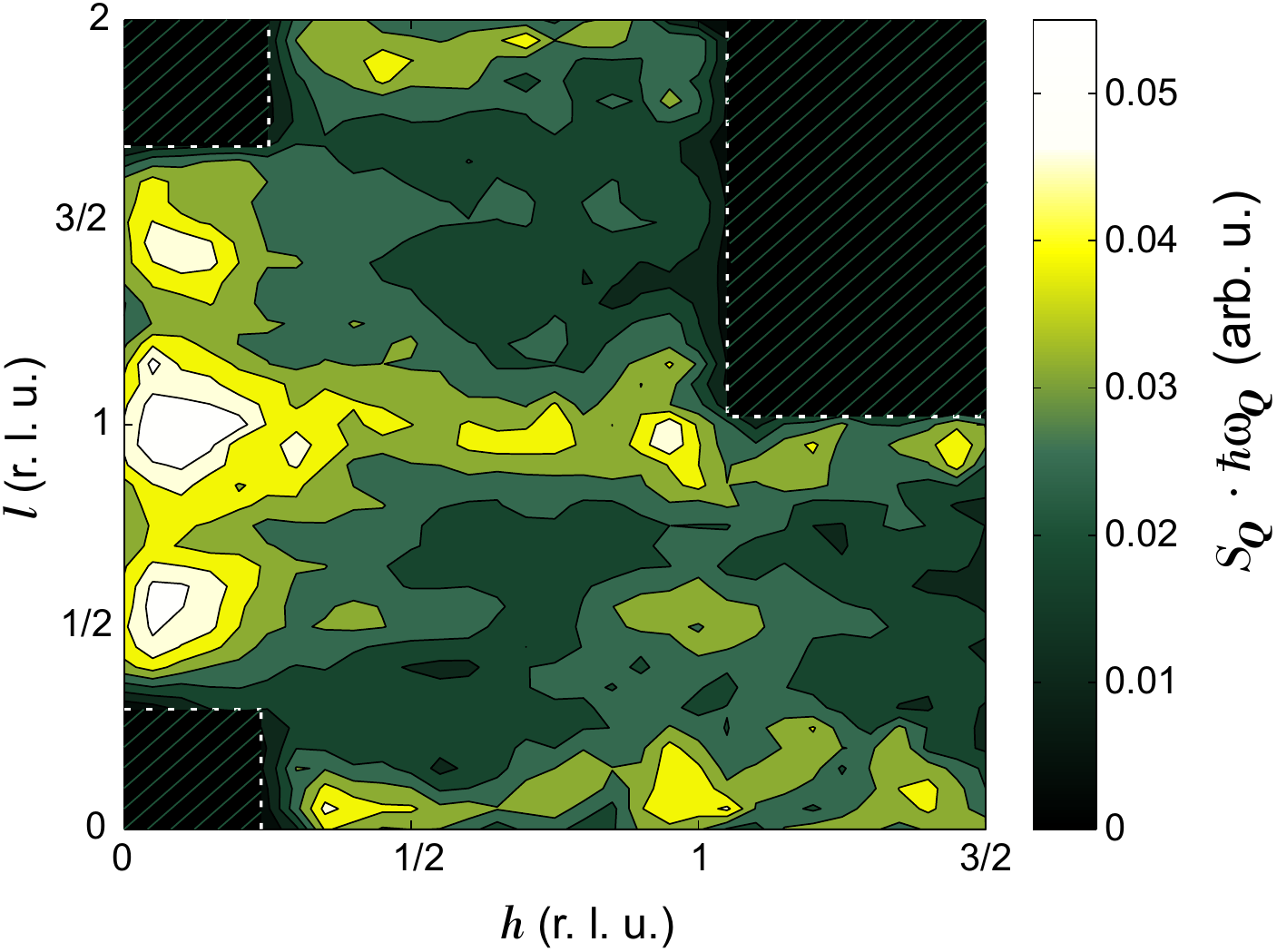}\\
  \caption{(Color online) Contour map of weighted peak intensity $\mathcal{S}_{\mathbf{Q}}\cdot\hbar\omega_{\mathbf{Q}}$ in the $(h,h,l)$ plane ($E_{\text{i}}=2.26$~meV). Note the half-period modulation along $\mathbf{c}^{\ast}$, i. e. the peaks at both integer and half-integer values of $l$.}\label{FIG:M1}
\end{figure}

Here we describe some features in our inelastic data that point to
the presence of additional magnetic interactions, not included in
the Hamiltonian~(\ref{EQ:Hamiltonian}). The first additional
interaction is the so-called ``diagonal'' Heisenberg exchange
between the two tetragonal sublattices. The presence of this
exchange $J_{d}$ was already noted in the ESR experiment by Zvyagin
\emph{et al.}\cite{Zvyagin_PRB_208_ESRinDTN} and confirmed in
neutron scattering investigation by Tsirulin \emph{et
al.}\cite{Tsyrulin_JPCM_2013_DTNneutrons} The introduction of
$J_{d}$ leads to additional double-period modulation of the
dispersion relation. This new periodicity can be indeed spotted in
the data. Figure~\ref{FIG:hw2} shows a comparison of dispersion
curves (actually, dispersion squared) measured along the $(h,h,0)$
and $(h,h,1)$ directions (symbols).  The difference between the
otherwise equivalent reciprocal-space directions is naturally
explained by a  non-zero $J_{d}$, as shown by the solid line which
corresponds to the best-fit value of $J_d$. For comparison, the
dashed line is the proposed DTN dispersion from
Ref.~\onlinecite{Zapf_PRL_2006_BECinDTN}, where $J_d$ is not
included.

The second additional term  that we consider is the next-nearest
neighbor exchange coupling $J_{c2}$ along the chain direction. The
evidence for this term comes from the very specific pattern found in
the intensity distribution. The product
$\mathcal{S}_{\mathbf{Q}}\cdot\hbar\omega_{\mathbf{Q}}$ shown in
Fig.~\ref{FIG:M1} exhibits a half-period modulation along the
$\mathbf{c}^{\ast}$ direction. This implies the presence of
additional interaction with the translation vector
$\mathbf{r}=2\mathbf{c}$. Additionally, $J_{2c}$ manifests itself in
a slight ``blunting'' of the $\cos$-like dispersion along the $c$
axis, near the top of the band. That, too, is consistent with our
data. Note that next-nearest-neighbor interactions along the $c$
axis could not be detected in the high-field experiment of
Ref.~\onlinecite{Tsyrulin_JPCM_2013_DTNneutrons}, as their
scattering plane was orthogonal to $\mathbf{c}$. Analyzing the
measured dispersion in DTNX, we find $J_{c2}$ to be
\textit{ferromagnetic}, and therefore nonfrustrating. Both $J_{d}$
and $J_{c2}$ corrections are relatively small, and don't
qualitatively affect the general picture for DTN.

\section{Self-consistent corrections to the RPA}\label{APP:RPArenorm}

\begin{table*}
  \centering
  \begin{tabular}{ l c l c l c l c l c l c l }
  \hline
  \hline
       & & \multicolumn{3}{c}{DTN} & & \multicolumn{7}{c}{6\% DTNX}\\
   \cline{3-5}\cline{7-13}
   & & \begin{tabular}{@{}c@{}}Zapf \\ \emph{et al.} (2006)\cite{Zapf_PRL_2006_BECinDTN}\end{tabular} & & \begin{tabular}{@{}c@{}} Tsirulin \\ \emph{et al.} (2013)\cite{Tsyrulin_JPCM_2013_DTNneutrons}\end{tabular} & & \begin{tabular}{@{}c@{}}Present study\\ Eq.~(\ref{EQ:Hamiltonian}) only\end{tabular} & & \begin{tabular}{@{}c@{}}Present study\\Eq.~(\ref{EQ:Hamiltonian})$+J_{d}$\end{tabular} & & \begin{tabular}{@{}c@{}}Present study \\ Eq.~(\ref{EQ:Hamiltonian})$+J_{c2}$\end{tabular} & & \begin{tabular}{@{}c@{}}Present study \\Eq.~(\ref{EQ:Hamiltonian})$+J_{d}+J_{c2}$\end{tabular}\\
   \hline
  $s^{2}$ & & $0.943$ & & $-$ & & $0.921$ & & $0.923$ & & $0.926$ & & $0.924$ \\
  $D$ & & $0.700(3)$~meV & & $0.767$~meV & & $0.682(3)$~meV & & $0.702(2)$~meV & & $0.702(2)$~meV & & $0.701(3)$~meV \\
  $J_{c}$ & & $0.150(3)$~meV & & $0.177$~meV & & $0.168(1)$~meV & & $0.171(1)$~meV & & $0.161(1)$~meV & & $0.162(1)$~meV \\
  $J_{a}$ & & $0.015(1)$~meV & & $0.0134(3)$~meV & & $0.0172(5)$~meV & & $0.0174(2)$~meV & & $0.0167(2)$~meV & & $0.0170(5)$~meV \\
  $J_{d}$ & & - & & $0.0069(9)$~meV & & - & & $0.0064(5)$~meV & & - & & $0.0065(4)$~meV \\
  $J_{c2}$ & & - & & - & & - & & - & & $-0.0103(7)$~meV & & $-0.0104(7)$~meV \\
  \hline
  $D/J_{c}$ & & $4.7(1)$ & & $4.34$ & & $4.06(3)$ & & $4.10(3)$ & & $ 4.36(3)$ & & $4.33(3)$ \\
  $J_{a}/J_{c}$ & & $0.098(7)$ & & $0.076(2)$ & & $0.102(2)$ & & $0.102(2)$ & & $0.104(2)$ & & $0.105(2)$ \\
  \hline
  \hline
\end{tabular}
  \caption{Results of fitting Eqs.~(\ref{EQ:RPAcorrected}-\ref{EQ:renormS2}) to our 6\% DTNX dispersion data and to those for the parent compound (Ref.~\onlinecite{Zapf_PRL_2006_BECinDTN}). For the parent compound, parameters deduced from the spin-wave dispersion measured in the saturated phase, are also listed. \cite{Tsyrulin_JPCM_2013_DTNneutrons}}\label{TAB:parametersR}
\end{table*}

\begin{table*}
  \centering
  \begin{tabular}{l c l c l c l c l c l}
  \hline\hline
     & & \multicolumn{1}{c}{DTN} & & \multicolumn{7}{c}{6\% DTNX}\\
   \cline{3-3}\cline{5-11}
    & & \begin{tabular}{@{}c@{}}Sizanov and \\ Syromyatnikov (2011)\cite{SizanovSyromyatnikov_PRB_2011_bosons4DTN}\end{tabular} & & \begin{tabular}{@{}c@{}}Present study\\Eq.~(\ref{EQ:Hamiltonian}) only\end{tabular} & & \begin{tabular}{@{}c@{}}Present study\\Eq.~(\ref{EQ:Hamiltonian})$+J_{d}$\end{tabular} & & \begin{tabular}{@{}c@{}}Present study \\Eq.~(\ref{EQ:Hamiltonian})$+J_{c2}$\end{tabular} & & \begin{tabular}{@{}c@{}}Present study \\ Eq.~(\ref{EQ:Hamiltonian})$+J_{d}+J_{c2}$\end{tabular}\\
   \hline
  $D$ & & $0.666$ meV & & $0.661(1)$ meV & & $0.661(1)$ meV & & $0.672(2)$ meV & & $0.672(2)$ meV \\
  $J_{c}$ & & $0.160$ meV & & $0.176(1)$ meV & & $0.177(1)$ meV & & $0.169(1)$ meV & & $170(2)$ meV \\
  $J_{a}$ & & $0.0172$ meV & & $0.0190(3)$ meV & & $0.0192(2)$ meV & & $0.0190(3)$ meV & & $0.0193(2)$ meV \\
  $J_{d}$ & & $0.0086$ meV & & - & & $0.0075(6)$ meV & & - & & $0.0075(5)$ meV \\
  $J_{c2}$ & & - & & - & & - & & $-0.007(1)$ meV & & $-0.007(1)$ meV \\
  \hline
    $D/J_{c}$ & & $4.16$ & & $3.75(3)$ & & $3.73(3)$ & & $3.96(4)$ & & $3.95(6)$ \\
  $J_{a}/J_{c}$ & & $0.108$ & & $0.108(2)$ & & $0.109(2)$ & & $0.112(2)$ & & $0.113(3)$ \\
  \hline\hline
\end{tabular}
  \caption{Results of fitting the dispersion obtained through the $1/D$ expansion~(\ref{EQ:dispersionSS}) to the data for 6\% DTNX. The results of an analogous treatment for
  the parent compound are also shown, following Ref.~\onlinecite{SizanovSyromyatnikov_PRB_2011_bosons4DTN}.}
  \label{TAB:parametersSS}
\end{table*}

The RPA result~(\ref{EQ:LinSW}) can be modified to account for the
quantum renormalization of the dispersion relation. In the framework
of a generalized spin-wave
approach,\cite{Zapf_PRL_2006_BECinDTN,ZhangWierschem_PRB_2013_DTNdispersion}
relying on artificial restriction of Hilbert space for the spin-wave
operators, Eq.~(\ref{EQ:LinSW}) is transformed into:

\begin{equation}\label{EQ:RPAcorrected}
\hbar\omega(\mathbf{Q})=\sqrt{D'^{2}+4D's^{2}\gamma(\mathbf{Q})}.
\end{equation}

Structurally this formula is identical to the RPA
result~(\ref{EQ:LinSW}). The difference is that the single-ion
anisotropy $D$ is replaced by the effective value $D'$ . In addition, instead
of the bare exchange couplings $J_{\mathbf{r}}$, one uses  renormalized
$s^{2}J_{\mathbf{r}}$. The values of $D'$ and
$s^{2}$ are related to the ``true'' values $D$ and $J_{\mathbf{r}}$
by a set of self-consistent equations:

\begin{equation}\label{EQ:renormD}
    D=D'\left(1+\dfrac{1}{4\pi^{3}}\int\limits_{\text{BZ}}\dfrac{\gamma(\mathbf{Q})}{\hbar\omega(\mathbf{Q})}d^{3}\mathbf{Q}\right),
\end{equation}

\begin{equation}\label{EQ:renormS2}
    s^{2}=2-\dfrac{1}{(2\pi)^{3}}\int\limits_{\text{BZ}}\dfrac{D'+2s^{2}\gamma(\mathbf{Q})}{\hbar\omega(\mathbf{Q})}d^{3}\mathbf{Q}.
\end{equation}

The integration is performed over the full Brillouin zone. Applying this correction to the parameters in
Table~\ref{TAB:parameters} results in the values, summarized in
Table~\ref{TAB:parametersR}. The parameters of pure DTN, determined
from the spin-wave dispersion in the fully polarized
phase,\cite{Tsyrulin_JPCM_2013_DTNneutrons} are also given for
comparison.

An alternative approach was developed by Sizanov and Syromyatnikov.\cite{SizanovSyromyatnikov_PRB_2011_bosons4DTN} Their dispersion equation, based upon the bosonization and subsequent expansion in $1/D$, is:


\begin{equation}\label{EQ:dispersionSS}
\begin{aligned}
\hbar\omega(\mathbf{Q})=&D+\frac{3}{2D}\int_{\text{BZ}}\frac{\gamma(\mathbf{P})^{2}}{(2\pi)^{3}}d^{3}\mathbf{P}\\
&+\frac{1}{D^{2}}\iint_{\text{BZ}}\frac{\gamma(\mathbf{P})\gamma(\mathbf{K})\gamma(\mathbf{P}-\mathbf{K})}{(2\pi)^{6}}d^{3}\mathbf{P}d^{3}\mathbf{K}\\
&+\gamma(\mathbf{Q})-\frac{1}{2D}\gamma(\mathbf{Q})^{2}+\frac{1}{2D^{2}}\gamma(\mathbf{Q})^{3}\\
&-\frac{1}{D^{2}}\gamma(\mathbf{Q})\int_{\text{BZ}}\frac{\frac{7}{4}\gamma(\mathbf{P})^{2}+\frac{1}{2}\gamma(\mathbf{P})\gamma(\mathbf{P}-\mathbf{Q})}{(2\pi)^{3}}d^{3}\mathbf{P}\\
&+\frac{5}{4D^{2}}\iint_{\text{BZ}}\frac{\gamma(\mathbf{P})\gamma(\mathbf{K})\gamma(\mathbf{P}-\mathbf{K}+\mathbf{Q})}{(2\pi)^{6}}d^{3}\mathbf{P}d^{3}\mathbf{K}.\\
\end{aligned}
\end{equation}

Applying this approach to  our data results in the interaction
parameters, summarized in Table~\ref{TAB:parametersSS}. Values for
pure DTN, as obtained in
Ref.~\onlinecite{SizanovSyromyatnikov_PRB_2011_bosons4DTN} by
analyzing the data of Ref.~\onlinecite{Zapf_PRL_2006_BECinDTN}, are
also given for comparison.  Although this approach leads to slightly different parameter values, the conclusion still holds: DTNX
approaches QCP primarily due to the reduction of the ratio $D/J_{c}$.

\bibliography{The_Library}

\begin{thebibliography}{58}%
\makeatletter
\providecommand \@ifxundefined [1]{%
 \@ifx{#1\undefined}
}%
\providecommand \@ifnum [1]{%
 \ifnum #1\expandafter \@firstoftwo
 \else \expandafter \@secondoftwo
 \fi
}%
\providecommand \@ifx [1]{%
 \ifx #1\expandafter \@firstoftwo
 \else \expandafter \@secondoftwo
 \fi
}%
\providecommand \natexlab [1]{#1}%
\providecommand \enquote  [1]{``#1''}%
\providecommand \bibnamefont  [1]{#1}%
\providecommand \bibfnamefont [1]{#1}%
\providecommand \citenamefont [1]{#1}%
\providecommand \href@noop [0]{\@secondoftwo}%
\providecommand \href [0]{\begingroup \@sanitize@url \@href}%
\providecommand \@href[1]{\@@startlink{#1}\@@href}%
\providecommand \@@href[1]{\endgroup#1\@@endlink}%
\providecommand \@sanitize@url [0]{\catcode `\\12\catcode `\$12\catcode
  `\&12\catcode `\#12\catcode `\^12\catcode `\_12\catcode `\%12\relax}%
\providecommand \@@startlink[1]{}%
\providecommand \@@endlink[0]{}%
\providecommand \url  [0]{\begingroup\@sanitize@url \@url }%
\providecommand \@url [1]{\endgroup\@href {#1}{\urlprefix }}%
\providecommand \urlprefix  [0]{URL }%
\providecommand \Eprint [0]{\href }%
\providecommand \doibase [0]{http://dx.doi.org/}%
\providecommand \selectlanguage [0]{\@gobble}%
\providecommand \bibinfo  [0]{\@secondoftwo}%
\providecommand \bibfield  [0]{\@secondoftwo}%
\providecommand \translation [1]{[#1]}%
\providecommand \BibitemOpen [0]{}%
\providecommand \bibitemStop [0]{}%
\providecommand \bibitemNoStop [0]{.\EOS\space}%
\providecommand \EOS [0]{\spacefactor3000\relax}%
\providecommand \BibitemShut  [1]{\csname bibitem#1\endcsname}%
\let\auto@bib@innerbib\@empty
\bibitem [{\citenamefont {Glarum}\ \emph {et~al.}(1991)\citenamefont {Glarum},
  \citenamefont {Geschwind}, \citenamefont {Lee}, \citenamefont {Kaplan},\ and\
  \citenamefont {Michel}}]{GlarumGeschwind_PRL_1991_HaldaneDoped}%
  \BibitemOpen
  \bibfield  {author} {\bibinfo {author} {\bibfnamefont {S.~H.}\ \bibnamefont
  {Glarum}}, \bibinfo {author} {\bibfnamefont {S.}~\bibnamefont {Geschwind}},
  \bibinfo {author} {\bibfnamefont {K.~M.}\ \bibnamefont {Lee}}, \bibinfo
  {author} {\bibfnamefont {M.~L.}\ \bibnamefont {Kaplan}}, \ and\ \bibinfo
  {author} {\bibfnamefont {J.}~\bibnamefont {Michel}},\ }\bibfield  {title}
  {\enquote {\bibinfo {title} {{Observation of fractional spin \textit{S} =1/2
  on open ends of \textit{S} =1 linear antiferromagnetic chains: Nonmagnetic
  doping}},}\ }\href {\doibase 10.1103/PhysRevLett.67.1614} {\bibfield
  {journal} {\bibinfo  {journal} {Phys. Rev. Lett.}\ }\textbf {\bibinfo
  {volume} {67}},\ \bibinfo {pages} {1614} (\bibinfo {year}
  {1991})}\BibitemShut {NoStop}%
\bibitem [{\citenamefont {Fisher}(1994)}]{Fisher_PRB_1994_RandomSinglet}%
  \BibitemOpen
  \bibfield  {author} {\bibinfo {author} {\bibfnamefont {D.~S.}\ \bibnamefont
  {Fisher}},\ }\bibfield  {title} {\enquote {\bibinfo {title} {Random
  antiferromagnetic quantum spin chains},}\ }\href {\doibase
  10.1103/PhysRevB.50.3799} {\bibfield  {journal} {\bibinfo  {journal} {Phys.
  Rev. B}\ }\textbf {\bibinfo {volume} {50}},\ \bibinfo {pages} {3799}
  (\bibinfo {year} {1994})}\BibitemShut {NoStop}%
\bibitem [{\citenamefont {Vojta}(2006)}]{Vojta_JPA_2006_DisorderReview}%
  \BibitemOpen
  \bibfield  {author} {\bibinfo {author} {\bibfnamefont {T.}~\bibnamefont
  {Vojta}},\ }\bibfield  {title} {\enquote {\bibinfo {title} {Rare region
  effects at classical, quantum and nonequilibrium phase transitions},}\ }\href
  {http://iopscience.iop.org/0305-4470/39/22/R01/} {\bibfield  {journal}
  {\bibinfo  {journal} {J. Phys. A}\ }\textbf {\bibinfo {volume} {39}},\
  \bibinfo {pages} {R143} (\bibinfo {year} {2006})}\BibitemShut {NoStop}%
\bibitem [{\citenamefont {Zheludev}\ and\ \citenamefont
  {Roscilde}(2013)}]{ZheludevRoscilde_CRPhysique_2013_ReviewDirtyBosons}%
  \BibitemOpen
  \bibfield  {author} {\bibinfo {author} {\bibfnamefont {A.}~\bibnamefont
  {Zheludev}}\ and\ \bibinfo {author} {\bibfnamefont {T.}~\bibnamefont
  {Roscilde}},\ }\bibfield  {title} {\enquote {\bibinfo {title} {Dirty-boson
  physics with magnetic insulators},}\ }\href {\doibase
  10.1016/j.crhy.2013.10.001} {\bibfield  {journal} {\bibinfo  {journal} {C. R.
  Physique}\ }\textbf {\bibinfo {volume} {14}},\ \bibinfo {pages} {740}
  (\bibinfo {year} {2013})}\BibitemShut {NoStop}%
\bibitem [{\citenamefont {Shender}\ and\ \citenamefont
  {Kivelson}(1991)}]{ShenderKivelson_PRL_1991_OrderSiteDisorder}%
  \BibitemOpen
  \bibfield  {author} {\bibinfo {author} {\bibfnamefont {E.~F.}\ \bibnamefont
  {Shender}}\ and\ \bibinfo {author} {\bibfnamefont {S.~A.}\ \bibnamefont
  {Kivelson}},\ }\bibfield  {title} {\enquote {\bibinfo {title}
  {Dilution-induced order in quasi-one-dimensional quantum antiferromagnets},}\
  }\href {\doibase 10.1103/PhysRevLett.66.2384} {\bibfield  {journal} {\bibinfo
   {journal} {Phys. Rev. Lett.}\ }\textbf {\bibinfo {volume} {66}},\ \bibinfo
  {pages} {2384} (\bibinfo {year} {1991})}\BibitemShut {NoStop}%
\bibitem [{\citenamefont {Uchiyama}\ \emph {et~al.}(1999)\citenamefont
  {Uchiyama}, \citenamefont {Sasago}, \citenamefont {Tsukada}, \citenamefont
  {Uchinokura}, \citenamefont {Zheludev}, \citenamefont {Hayashi},
  \citenamefont {Miura},\ and\ \citenamefont
  {B\"oni}}]{Uchiyama_PRL_1999_PNVOorder}%
  \BibitemOpen
  \bibfield  {author} {\bibinfo {author} {\bibfnamefont {Y.}~\bibnamefont
  {Uchiyama}}, \bibinfo {author} {\bibfnamefont {Y.}~\bibnamefont {Sasago}},
  \bibinfo {author} {\bibfnamefont {I.}~\bibnamefont {Tsukada}}, \bibinfo
  {author} {\bibfnamefont {K.}~\bibnamefont {Uchinokura}}, \bibinfo {author}
  {\bibfnamefont {A.}~\bibnamefont {Zheludev}}, \bibinfo {author}
  {\bibfnamefont {T.}~\bibnamefont {Hayashi}}, \bibinfo {author} {\bibfnamefont
  {N.}~\bibnamefont {Miura}}, \ and\ \bibinfo {author} {\bibfnamefont
  {P.}~\bibnamefont {B\"oni}},\ }\bibfield  {title} {\enquote {\bibinfo {title}
  {{Spin-Vacancy-Induced Long-Range Order in a New Haldane-Gap
  Antiferromagnet}},}\ }\href {\doibase 10.1103/PhysRevLett.83.632} {\bibfield
  {journal} {\bibinfo  {journal} {Phys. Rev. Lett.}\ }\textbf {\bibinfo
  {volume} {83}},\ \bibinfo {pages} {632} (\bibinfo {year} {1999})}\BibitemShut
  {NoStop}%
\bibitem [{\citenamefont
  {Uchinokura}(2002)}]{Uchinokura_JPCM_2002_CuGeO3review}%
  \BibitemOpen
  \bibfield  {author} {\bibinfo {author} {\bibfnamefont {K.}~\bibnamefont
  {Uchinokura}},\ }\bibfield  {title} {\enquote {\bibinfo {title}
  {{Spin-Peierls transition in CuGeO$_3$ and impurity-induced ordered phases in
  low-dimensional spin-gap systems}},}\ }\href {\doibase
  10.1088/0953-8984/14/10/201} {\bibfield  {journal} {\bibinfo  {journal} {J.
  Phys.: Condens. Matter}\ }\textbf {\bibinfo {volume} {14}},\ \bibinfo {pages}
  {R195} (\bibinfo {year} {2002})}\BibitemShut {NoStop}%
\bibitem [{\citenamefont {Smirnov}\ and\ \citenamefont
  {Glazkov}(2007)}]{SmirnovGlazkov_JETP_2007_disorderreview}%
  \BibitemOpen
  \bibfield  {author} {\bibinfo {author} {\bibfnamefont {A.~I.}\ \bibnamefont
  {Smirnov}}\ and\ \bibinfo {author} {\bibfnamefont {V.~N.}\ \bibnamefont
  {Glazkov}},\ }\bibfield  {title} {\enquote {\bibinfo {title} {{Mesoscopic
  spin clusters, phase separation, and induced order in spin-gap magnets: A
  review}},}\ }\href@noop {} {\bibfield  {journal} {\bibinfo  {journal} {J.
  Exp. Theor. Phys.}\ }\textbf {\bibinfo {volume} {105}},\ \bibinfo {pages}
  {861} (\bibinfo {year} {2007})}\BibitemShut {NoStop}%
\bibitem [{\citenamefont {Sachdev}(2011)}]{Sachdev_2011_QPTBook}%
  \BibitemOpen
  \bibfield  {author} {\bibinfo {author} {\bibfnamefont {S.}~\bibnamefont
  {Sachdev}},\ }\href
  {http://www.cambridge.org/ch/academic/subjects/physics/condensed-matter-physics-nanoscience-and-mesoscopic-physics/quantum-phase-transitions-2nd-edition?format=HB}
  {\emph {\bibinfo {title} {Quantum Phase Transitions}}}\ (\bibinfo
  {publisher} {Cambridge University Press, U.K.},\ \bibinfo {year}
  {2011})\BibitemShut {NoStop}%
\bibitem [{\citenamefont {Sachdev}(2000)}]{Sachdev_Science_2000_QCPs}%
  \BibitemOpen
  \bibfield  {author} {\bibinfo {author} {\bibfnamefont {S.}~\bibnamefont
  {Sachdev}},\ }\bibfield  {title} {\enquote {\bibinfo {title} {Quantum
  criticality: Competing ground states in low dimensions},}\ }\href {\doibase
  10.1126/science.288.5465.475} {\bibfield  {journal} {\bibinfo  {journal}
  {Science}\ }\textbf {\bibinfo {volume} {288}},\ \bibinfo {pages} {475}
  (\bibinfo {year} {2000})}\BibitemShut {NoStop}%
\bibitem [{\citenamefont {Sachdev}(2008)}]{Sachdev_NPhys_2008_QCPs}%
  \BibitemOpen
  \bibfield  {author} {\bibinfo {author} {\bibfnamefont {S.}~\bibnamefont
  {Sachdev}},\ }\bibfield  {title} {\enquote {\bibinfo {title} {Quantum
  magnetism and criticality},}\ }\href {\doibase 10.1038/nphys894} {\bibfield
  {journal} {\bibinfo  {journal} {Nat. Physics}\ }\textbf {\bibinfo {volume}
  {4}},\ \bibinfo {pages} {173} (\bibinfo {year} {2008})}\BibitemShut {NoStop}%
\bibitem [{\citenamefont {Vojta}(2013)}]{Vojta_PRL_2013_InGap}%
  \BibitemOpen
  \bibfield  {author} {\bibinfo {author} {\bibfnamefont {M.}~\bibnamefont
  {Vojta}},\ }\bibfield  {title} {\enquote {\bibinfo {title} {Excitation
  spectra of disordered dimer magnets near quantum criticality},}\ }\href
  {\doibase 10.1103/PhysRevLett.111.097202} {\bibfield  {journal} {\bibinfo
  {journal} {Phys. Rev. Lett.}\ }\textbf {\bibinfo {volume} {111}},\ \bibinfo
  {pages} {097202} (\bibinfo {year} {2013})}\BibitemShut {NoStop}%
\bibitem [{Note1()}]{Note1}%
  \BibitemOpen
  \bibinfo {note} {{This $z=1$ QCP is not to be confused with the $z=2$
  field-induced transition alike Bose--Einstein condensation of
  magnons.}}\BibitemShut {Stop}%
\bibitem [{Note2()}]{Note2}%
  \BibitemOpen
  \bibinfo {note} {This is a coexistence of sharp Bragg peaks with extremely
  broadened excitations at finite energies.}\BibitemShut {Stop}%
\bibitem [{\citenamefont {Landee}\ and\ \citenamefont
  {Turnbull}(2013)}]{LandeeTurnbull_EuJInChem_2013_ReviewMagnets}%
  \BibitemOpen
  \bibfield  {author} {\bibinfo {author} {\bibfnamefont {C.~P.}\ \bibnamefont
  {Landee}}\ and\ \bibinfo {author} {\bibfnamefont {M.~M.}\ \bibnamefont
  {Turnbull}},\ }\bibfield  {title} {\enquote {\bibinfo {title} {{Recent
  Developments in Low-Dimensional Copper(II) Molecular Magnets}},}\ }\href
  {\doibase 10.1002/ejic.201300133} {\bibfield  {journal} {\bibinfo  {journal}
  {Eur. J. Inorg. Chem.}\ }\textbf {\bibinfo {volume} {2013}},\ \bibinfo
  {pages} {2266} (\bibinfo {year} {2013})}\BibitemShut {NoStop}%
\bibitem [{\citenamefont {Yankova}\ \emph {et~al.}(2012)\citenamefont
  {Yankova}, \citenamefont {H\"{u}vonen}, \citenamefont {M\"{u}hlbauer},
  \citenamefont {Schmidiger}, \citenamefont {Wulf}, \citenamefont {Zhao},
  \citenamefont {Zheludev}, \citenamefont {Hong}, \citenamefont {Garlea},
  \citenamefont {Custelcean},\ and\ \citenamefont
  {Ehlers}}]{Yankova_PhilMag_2012_ReviewXtals}%
  \BibitemOpen
  \bibfield  {author} {\bibinfo {author} {\bibfnamefont {T.}~\bibnamefont
  {Yankova}}, \bibinfo {author} {\bibfnamefont {D.}~\bibnamefont
  {H\"{u}vonen}}, \bibinfo {author} {\bibfnamefont {S.}~\bibnamefont
  {M\"{u}hlbauer}}, \bibinfo {author} {\bibfnamefont {D.}~\bibnamefont
  {Schmidiger}}, \bibinfo {author} {\bibfnamefont {E.}~\bibnamefont {Wulf}},
  \bibinfo {author} {\bibfnamefont {S.}~\bibnamefont {Zhao}}, \bibinfo {author}
  {\bibfnamefont {A.}~\bibnamefont {Zheludev}}, \bibinfo {author}
  {\bibfnamefont {T.}~\bibnamefont {Hong}}, \bibinfo {author} {\bibfnamefont
  {V.~O.}\ \bibnamefont {Garlea}}, \bibinfo {author} {\bibfnamefont
  {R.}~\bibnamefont {Custelcean}}, \ and\ \bibinfo {author} {\bibfnamefont
  {G.}~\bibnamefont {Ehlers}},\ }\bibfield  {title} {\enquote {\bibinfo {title}
  {Crystals for neutron scattering studies of quantum magnetism},}\ }\href
  {\doibase 10.1080/14786435.2012.669072} {\bibfield  {journal} {\bibinfo
  {journal} {Philos. Mag.}\ }\textbf {\bibinfo {volume} {92}},\ \bibinfo
  {pages} {2629} (\bibinfo {year} {2012})}\BibitemShut {NoStop}%
\bibitem [{\citenamefont {Oosawa}\ and\ \citenamefont
  {Tanaka}(2002)}]{OosawaTanaka_PRB_2002_TlCuCl3disorder}%
  \BibitemOpen
  \bibfield  {author} {\bibinfo {author} {\bibfnamefont {A.}~\bibnamefont
  {Oosawa}}\ and\ \bibinfo {author} {\bibfnamefont {H.}~\bibnamefont
  {Tanaka}},\ }\bibfield  {title} {\enquote {\bibinfo {title} {{Random bond
  effect in the quantum spin system
  $({\mathrm{Tl}}_{1-x}{\mathrm{K}}_{x}){\mathrm{CuCl}}_{3}$}},}\ }\href
  {\doibase 10.1103/PhysRevB.65.184437} {\bibfield  {journal} {\bibinfo
  {journal} {Phys. Rev. B}\ }\textbf {\bibinfo {volume} {65}},\ \bibinfo
  {pages} {184437} (\bibinfo {year} {2002})}\BibitemShut {NoStop}%
\bibitem [{\citenamefont {N\'afr\'adi}\ \emph {et~al.}(2013)\citenamefont
  {N\'afr\'adi}, \citenamefont {Keller}, \citenamefont {Manaka}, \citenamefont
  {Stuhr}, \citenamefont {Zheludev},\ and\ \citenamefont
  {Keimer}}]{NafradiKeller_PRB_2013_IPAXblueshift}%
  \BibitemOpen
  \bibfield  {author} {\bibinfo {author} {\bibfnamefont {B.}~\bibnamefont
  {N\'afr\'adi}}, \bibinfo {author} {\bibfnamefont {T.}~\bibnamefont {Keller}},
  \bibinfo {author} {\bibfnamefont {H.}~\bibnamefont {Manaka}}, \bibinfo
  {author} {\bibfnamefont {U.}~\bibnamefont {Stuhr}}, \bibinfo {author}
  {\bibfnamefont {A.}~\bibnamefont {Zheludev}}, \ and\ \bibinfo {author}
  {\bibfnamefont {B.}~\bibnamefont {Keimer}},\ }\bibfield  {title} {\enquote
  {\bibinfo {title} {{Bond randomness induced magnon decoherence in a
  spin-$\frac{1}{2}$ ladder compound}},}\ }\href {\doibase
  10.1103/PhysRevB.87.020408} {\bibfield  {journal} {\bibinfo  {journal} {Phys.
  Rev. B}\ }\textbf {\bibinfo {volume} {87}},\ \bibinfo {pages} {020408}
  (\bibinfo {year} {2013})}\BibitemShut {NoStop}%
\bibitem [{\citenamefont {H\"uvonen}\ \emph {et~al.}(2012)\citenamefont
  {H\"uvonen}, \citenamefont {Zhao}, \citenamefont {M\aa{}nsson}, \citenamefont
  {Yankova}, \citenamefont {Ressouche}, \citenamefont {Niedermayer},
  \citenamefont {Laver}, \citenamefont {Gvasaliya},\ and\ \citenamefont
  {Zheludev}}]{Huvonen_PRB_2012_PHCXdiffraction}%
  \BibitemOpen
  \bibfield  {author} {\bibinfo {author} {\bibfnamefont {D.}~\bibnamefont
  {H\"uvonen}}, \bibinfo {author} {\bibfnamefont {S.}~\bibnamefont {Zhao}},
  \bibinfo {author} {\bibfnamefont {M.}~\bibnamefont {M\aa{}nsson}}, \bibinfo
  {author} {\bibfnamefont {T.}~\bibnamefont {Yankova}}, \bibinfo {author}
  {\bibfnamefont {E.}~\bibnamefont {Ressouche}}, \bibinfo {author}
  {\bibfnamefont {C.}~\bibnamefont {Niedermayer}}, \bibinfo {author}
  {\bibfnamefont {M.}~\bibnamefont {Laver}}, \bibinfo {author} {\bibfnamefont
  {S.~N.}\ \bibnamefont {Gvasaliya}}, \ and\ \bibinfo {author} {\bibfnamefont
  {A.}~\bibnamefont {Zheludev}},\ }\bibfield  {title} {\enquote {\bibinfo
  {title} {Field-induced criticality in a gapped quantum magnet with bond
  disorder},}\ }\href {\doibase 10.1103/PhysRevB.85.100410} {\bibfield
  {journal} {\bibinfo  {journal} {Phys. Rev. B}\ }\textbf {\bibinfo {volume}
  {85}},\ \bibinfo {pages} {100410} (\bibinfo {year} {2012})}\BibitemShut
  {NoStop}%
\bibitem [{\citenamefont {H\"{u}vonen}\ \emph {et~al.}(2012)\citenamefont
  {H\"{u}vonen}, \citenamefont {Zhao}, \citenamefont {Ehlers}, \citenamefont
  {M\aa{}nsson}, \citenamefont {Gvasaliya},\ and\ \citenamefont
  {Zheludev}}]{Huvonen_PRB_2012_PHCXneutron}%
  \BibitemOpen
  \bibfield  {author} {\bibinfo {author} {\bibfnamefont {D.}~\bibnamefont
  {H\"{u}vonen}}, \bibinfo {author} {\bibfnamefont {S.}~\bibnamefont {Zhao}},
  \bibinfo {author} {\bibfnamefont {G.}~\bibnamefont {Ehlers}}, \bibinfo
  {author} {\bibfnamefont {M.}~\bibnamefont {M\aa{}nsson}}, \bibinfo {author}
  {\bibfnamefont {S.~N.}\ \bibnamefont {Gvasaliya}}, \ and\ \bibinfo {author}
  {\bibfnamefont {A.}~\bibnamefont {Zheludev}},\ }\bibfield  {title} {\enquote
  {\bibinfo {title} {Excitations in a quantum spin liquid with random bonds},}\
  }\href {\doibase 10.1103/PhysRevB.86.214408} {\bibfield  {journal} {\bibinfo
  {journal} {Phys. Rev. B}\ }\textbf {\bibinfo {volume} {86}},\ \bibinfo
  {pages} {214408} (\bibinfo {year} {2012})}\BibitemShut {NoStop}%
\bibitem [{\citenamefont {H\"{u}vonen}\ \emph {et~al.}(2013)\citenamefont
  {H\"{u}vonen}, \citenamefont {Ballon},\ and\ \citenamefont
  {Zheludev}}]{Huvonen_PRB_2013_PHCXphasediagram}%
  \BibitemOpen
  \bibfield  {author} {\bibinfo {author} {\bibfnamefont {D.}~\bibnamefont
  {H\"{u}vonen}}, \bibinfo {author} {\bibfnamefont {G.}~\bibnamefont {Ballon}},
  \ and\ \bibinfo {author} {\bibfnamefont {A.}~\bibnamefont {Zheludev}},\
  }\bibfield  {title} {\enquote {\bibinfo {title} {Field-concentration phase
  diagram of a quantum spin liquid with bond defects},}\ }\href {\doibase
  10.1103/PhysRevB.88.094402} {\bibfield  {journal} {\bibinfo  {journal} {Phys.
  Rev. B}\ }\textbf {\bibinfo {volume} {88}},\ \bibinfo {pages} {094402}
  (\bibinfo {year} {2013})}\BibitemShut {NoStop}%
\bibitem [{\citenamefont {Glazkov}\ \emph {et~al.}(2014)\citenamefont
  {Glazkov}, \citenamefont {Skoblin}, \citenamefont {H\"{u}vonen},
  \citenamefont {Yankova},\ and\ \citenamefont
  {Zheludev}}]{Glazkov_JPCM_2014_PHCXesr}%
  \BibitemOpen
  \bibfield  {author} {\bibinfo {author} {\bibfnamefont {V.~N.}\ \bibnamefont
  {Glazkov}}, \bibinfo {author} {\bibfnamefont {G.}~\bibnamefont {Skoblin}},
  \bibinfo {author} {\bibfnamefont {D.}~\bibnamefont {H\"{u}vonen}}, \bibinfo
  {author} {\bibfnamefont {T.~S.}\ \bibnamefont {Yankova}}, \ and\ \bibinfo
  {author} {\bibfnamefont {A}~\bibnamefont {Zheludev}},\ }\bibfield  {title}
  {\enquote {\bibinfo {title} {{Formation of gapless triplets in the bond-doped
  spin-gap antiferromagnet (C$_4$H$_{12}$N$_2$)(Cu$_2$Cl$_6$)}},}\ }\href
  {\doibase 10.1088/0953-8984/26/48/486002} {\bibfield  {journal} {\bibinfo
  {journal} {J. Phys.: Condens. Matter}\ }\textbf {\bibinfo {volume} {26}},\
  \bibinfo {pages} {486002} (\bibinfo {year} {2014})}\BibitemShut {NoStop}%
\bibitem [{\citenamefont {Wulf}\ \emph {et~al.}(2011)\citenamefont {Wulf},
  \citenamefont {M\"uhlbauer}, \citenamefont {Yankova},\ and\ \citenamefont
  {Zheludev}}]{WulfMuhlbauer_PRB_2011_SulDisordered}%
  \BibitemOpen
  \bibfield  {author} {\bibinfo {author} {\bibfnamefont {E.}~\bibnamefont
  {Wulf}}, \bibinfo {author} {\bibfnamefont {S.}~\bibnamefont {M\"uhlbauer}},
  \bibinfo {author} {\bibfnamefont {T.}~\bibnamefont {Yankova}}, \ and\
  \bibinfo {author} {\bibfnamefont {A.}~\bibnamefont {Zheludev}},\ }\bibfield
  {title} {\enquote {\bibinfo {title} {Disorder instability of the magnon
  condensate in a frustrated spin ladder},}\ }\href {\doibase
  10.1103/PhysRevB.84.174414} {\bibfield  {journal} {\bibinfo  {journal} {Phys.
  Rev. B}\ }\textbf {\bibinfo {volume} {84}},\ \bibinfo {pages} {174414}
  (\bibinfo {year} {2011})}\BibitemShut {NoStop}%
\bibitem [{\citenamefont {Keith}\ \emph {et~al.}(2011)\citenamefont {Keith},
  \citenamefont {Xiao}, \citenamefont {Landee}, \citenamefont {Turnbull},\ and\
  \citenamefont {Zheludev}}]{Keith_Polyhedron_2011_CQX}%
  \BibitemOpen
  \bibfield  {author} {\bibinfo {author} {\bibfnamefont {B.~C.}\ \bibnamefont
  {Keith}}, \bibinfo {author} {\bibfnamefont {F.}~\bibnamefont {Xiao}},
  \bibinfo {author} {\bibfnamefont {C.~P.}\ \bibnamefont {Landee}}, \bibinfo
  {author} {\bibfnamefont {M.~M.}\ \bibnamefont {Turnbull}}, \ and\ \bibinfo
  {author} {\bibfnamefont {A.}~\bibnamefont {Zheludev}},\ }\bibfield  {title}
  {\enquote {\bibinfo {title} {{Random exchange in the spin ladder
  Cu(quinoxaline)X$_2$ (X=Cl, Br)}},}\ }\href {\doibase
  10.1016/j.poly.2011.02.016} {\bibfield  {journal} {\bibinfo  {journal}
  {Polyhedron}\ }\textbf {\bibinfo {volume} {30}},\ \bibinfo {pages} {3006}
  (\bibinfo {year} {2011})}\BibitemShut {NoStop}%
\bibitem [{\citenamefont {Povarov}\ \emph {et~al.}(2014)\citenamefont
  {Povarov}, \citenamefont {Lorenz}, \citenamefont {Xiao}, \citenamefont
  {Landee}, \citenamefont {Krasnikova},\ and\ \citenamefont
  {Zheludev}}]{PovarovLorenz_JMMM_2014_CQX}%
  \BibitemOpen
  \bibfield  {author} {\bibinfo {author} {\bibfnamefont {\relax{K. Yu.}}\
  \bibnamefont {Povarov}}, \bibinfo {author} {\bibfnamefont {W.~E.~A.}\
  \bibnamefont {Lorenz}}, \bibinfo {author} {\bibfnamefont {F.}~\bibnamefont
  {Xiao}}, \bibinfo {author} {\bibfnamefont {C.~P.}\ \bibnamefont {Landee}},
  \bibinfo {author} {\bibfnamefont {Y.}~\bibnamefont {Krasnikova}}, \ and\
  \bibinfo {author} {\bibfnamefont {A.}~\bibnamefont {Zheludev}},\ }\bibfield
  {title} {\enquote {\bibinfo {title} {{The tunable quantum spin ladder
  Cu(Qnx)(Cl$_{(1-x)}$Br$_x$)$_2$}},}\ }\href {\doibase
  10.1016/j.jmmm.2014.06.060} {\bibfield  {journal} {\bibinfo  {journal} {J.
  Magn. Magn. Mater.}\ }\textbf {\bibinfo {volume} {370}},\ \bibinfo {pages}
  {62} (\bibinfo {year} {2014})}\BibitemShut {NoStop}%
\bibitem [{\citenamefont {Paduan-Filho}\ \emph {et~al.}(1981)\citenamefont
  {Paduan-Filho}, \citenamefont {Chirico}, \citenamefont {Joung},\ and\
  \citenamefont {Carlin}}]{PaduanFilho_JChemPhys_1981_DTNfirst}%
  \BibitemOpen
  \bibfield  {author} {\bibinfo {author} {\bibfnamefont {A.}~\bibnamefont
  {Paduan-Filho}}, \bibinfo {author} {\bibfnamefont {R.~D.}\ \bibnamefont
  {Chirico}}, \bibinfo {author} {\bibfnamefont {K.~O.}\ \bibnamefont {Joung}},
  \ and\ \bibinfo {author} {\bibfnamefont {R.~L.}\ \bibnamefont {Carlin}},\
  }\bibfield  {title} {\enquote {\bibinfo {title} {Field-induced magnetic
  ordering in uniaxial nickel systems: A second example},}\ }\href {\doibase
  http://dx.doi.org/10.1063/1.441589} {\bibfield  {journal} {\bibinfo
  {journal} {J. Chem. Phys.}\ }\textbf {\bibinfo {volume} {74}},\ \bibinfo
  {pages} {4103} (\bibinfo {year} {1981})}\BibitemShut {NoStop}%
\bibitem [{\citenamefont {Zapf}\ \emph {et~al.}(2006)\citenamefont {Zapf},
  \citenamefont {Zocco}, \citenamefont {Hansen}, \citenamefont {Jaime},
  \citenamefont {Harrison}, \citenamefont {Batista}, \citenamefont
  {Kenzelmann}, \citenamefont {Niedermayer}, \citenamefont {Lacerda},\ and\
  \citenamefont {Paduan-Filho}}]{Zapf_PRL_2006_BECinDTN}%
  \BibitemOpen
  \bibfield  {author} {\bibinfo {author} {\bibfnamefont {V.~S.}\ \bibnamefont
  {Zapf}}, \bibinfo {author} {\bibfnamefont {D.}~\bibnamefont {Zocco}},
  \bibinfo {author} {\bibfnamefont {B.~R.}\ \bibnamefont {Hansen}}, \bibinfo
  {author} {\bibfnamefont {M.}~\bibnamefont {Jaime}}, \bibinfo {author}
  {\bibfnamefont {N.}~\bibnamefont {Harrison}}, \bibinfo {author}
  {\bibfnamefont {C.~D.}\ \bibnamefont {Batista}}, \bibinfo {author}
  {\bibfnamefont {M.}~\bibnamefont {Kenzelmann}}, \bibinfo {author}
  {\bibfnamefont {C.}~\bibnamefont {Niedermayer}}, \bibinfo {author}
  {\bibfnamefont {A.}~\bibnamefont {Lacerda}}, \ and\ \bibinfo {author}
  {\bibfnamefont {A.}~\bibnamefont {Paduan-Filho}},\ }\bibfield  {title}
  {\enquote {\bibinfo {title} {{Bose-Einstein Condensation of $S=1$ Nickel Spin
  Degrees of Freedom in
  ${\mathrm{NiCl}}_{2}\mathrm{\text{-}}4\mathrm{SC}({\mathrm{NH}}_{2}{)}_{2}$}},}\
  }\href {\doibase 10.1103/PhysRevLett.96.077204} {\bibfield  {journal}
  {\bibinfo  {journal} {Phys. Rev. Lett.}\ }\textbf {\bibinfo {volume} {96}},\
  \bibinfo {pages} {077204} (\bibinfo {year} {2006})}\BibitemShut {NoStop}%
\bibitem [{\citenamefont {Zvyagin}\ \emph {et~al.}(2007)\citenamefont
  {Zvyagin}, \citenamefont {Wosnitza}, \citenamefont {Batista}, \citenamefont
  {Tsukamoto}, \citenamefont {Kawashima}, \citenamefont {Krzystek},
  \citenamefont {Zapf}, \citenamefont {Jaime}, \citenamefont {Oliveira},\ and\
  \citenamefont {Paduan-Filho}}]{Zvyagin_PRL_2007_ESRinDTN}%
  \BibitemOpen
  \bibfield  {author} {\bibinfo {author} {\bibfnamefont {S.~A.}\ \bibnamefont
  {Zvyagin}}, \bibinfo {author} {\bibfnamefont {J.}~\bibnamefont {Wosnitza}},
  \bibinfo {author} {\bibfnamefont {C.~D.}\ \bibnamefont {Batista}}, \bibinfo
  {author} {\bibfnamefont {M.}~\bibnamefont {Tsukamoto}}, \bibinfo {author}
  {\bibfnamefont {N.}~\bibnamefont {Kawashima}}, \bibinfo {author}
  {\bibfnamefont {J.}~\bibnamefont {Krzystek}}, \bibinfo {author}
  {\bibfnamefont {V.~S.}\ \bibnamefont {Zapf}}, \bibinfo {author}
  {\bibfnamefont {M.}~\bibnamefont {Jaime}}, \bibinfo {author} {\bibfnamefont
  {N.~F.}\ \bibnamefont {Oliveira}}, \ and\ \bibinfo {author} {\bibfnamefont
  {A.}~\bibnamefont {Paduan-Filho}},\ }\bibfield  {title} {\enquote {\bibinfo
  {title} {{Magnetic Excitations in the Spin-1 Anisotropic Heisenberg
  Antiferromagnetic Chain System
  ${\mathrm{NiCl}}_{2}\mathrm{\text{-}}4\mathrm{SC}({\mathrm{NH}}_{2}{)}_{2}$}},}\
  }\href {\doibase 10.1103/PhysRevLett.98.047205} {\bibfield  {journal}
  {\bibinfo  {journal} {Phys. Rev. Lett.}\ }\textbf {\bibinfo {volume} {98}},\
  \bibinfo {pages} {047205} (\bibinfo {year} {2007})}\BibitemShut {NoStop}%
\bibitem [{\citenamefont {Yin}\ \emph {et~al.}(2008)\citenamefont {Yin},
  \citenamefont {Xia}, \citenamefont {Zapf}, \citenamefont {Sullivan},\ and\
  \citenamefont {Paduan-Filho}}]{YinXia_PRL_2008_DTNcritical}%
  \BibitemOpen
  \bibfield  {author} {\bibinfo {author} {\bibfnamefont {L.}~\bibnamefont
  {Yin}}, \bibinfo {author} {\bibfnamefont {J.~S.}\ \bibnamefont {Xia}},
  \bibinfo {author} {\bibfnamefont {V.~S.}\ \bibnamefont {Zapf}}, \bibinfo
  {author} {\bibfnamefont {N.~S.}\ \bibnamefont {Sullivan}}, \ and\ \bibinfo
  {author} {\bibfnamefont {A.}~\bibnamefont {Paduan-Filho}},\ }\bibfield
  {title} {\enquote {\bibinfo {title} {{Direct Measurement of the Bose-Einstein
  Condensation Universality Class in
  ${\mathrm{NiCl}}_{2}\mathrm{\text{-}}4\mathrm{SC}({\mathrm{NH}}_{2}{)}_{2}$
  at Ultralow Temperatures}},}\ }\href {\doibase
  10.1103/PhysRevLett.101.187205} {\bibfield  {journal} {\bibinfo  {journal}
  {Phys. Rev. Lett.}\ }\textbf {\bibinfo {volume} {101}},\ \bibinfo {pages}
  {187205} (\bibinfo {year} {2008})}\BibitemShut {NoStop}%
\bibitem [{\citenamefont {Mukhopadhyay}\ \emph {et~al.}(2012)\citenamefont
  {Mukhopadhyay}, \citenamefont {Klanj\v{s}ek}, \citenamefont {Grbi\'{c}},
  \citenamefont {Blinder}, \citenamefont {Mayaffre}, \citenamefont {Berthier},
  \citenamefont {Horvati\'{c}}, \citenamefont {Continentino}, \citenamefont
  {Paduan-Filho}, \citenamefont {Chiari},\ and\ \citenamefont
  {Piovesana}}]{Mukhopadhyay_PRL_2012_NMRinDTNandDIMPY}%
  \BibitemOpen
  \bibfield  {author} {\bibinfo {author} {\bibfnamefont {S.}~\bibnamefont
  {Mukhopadhyay}}, \bibinfo {author} {\bibfnamefont {M.}~\bibnamefont
  {Klanj\v{s}ek}}, \bibinfo {author} {\bibfnamefont {M.~S.}\ \bibnamefont
  {Grbi\'{c}}}, \bibinfo {author} {\bibfnamefont {R.}~\bibnamefont {Blinder}},
  \bibinfo {author} {\bibfnamefont {H.}~\bibnamefont {Mayaffre}}, \bibinfo
  {author} {\bibfnamefont {C.}~\bibnamefont {Berthier}}, \bibinfo {author}
  {\bibfnamefont {M.}~\bibnamefont {Horvati\'{c}}}, \bibinfo {author}
  {\bibfnamefont {M.~A.}\ \bibnamefont {Continentino}}, \bibinfo {author}
  {\bibfnamefont {A.}~\bibnamefont {Paduan-Filho}}, \bibinfo {author}
  {\bibfnamefont {B.}~\bibnamefont {Chiari}}, \ and\ \bibinfo {author}
  {\bibfnamefont {O.}~\bibnamefont {Piovesana}},\ }\bibfield  {title} {\enquote
  {\bibinfo {title} {Quantum-critical spin dynamics in quasi-one-dimensional
  antiferromagnets},}\ }\href {\doibase 10.1103/PhysRevLett.109.177206}
  {\bibfield  {journal} {\bibinfo  {journal} {Phys. Rev. Lett.}\ }\textbf
  {\bibinfo {volume} {109}},\ \bibinfo {pages} {177206} (\bibinfo {year}
  {2012})}\BibitemShut {NoStop}%
\bibitem [{\citenamefont {Tsyrulin}\ \emph {et~al.}(2013)\citenamefont
  {Tsyrulin}, \citenamefont {Batista}, \citenamefont {Zapf}, \citenamefont
  {Jaime}, \citenamefont {Hansen}, \citenamefont {Niedermayer}, \citenamefont
  {Rule}, \citenamefont {Habicht}, \citenamefont {Prokes}, \citenamefont
  {Kiefer}, \citenamefont {Ressouche}, \citenamefont {Paduan-Filho},\ and\
  \citenamefont {Kenzelmann}}]{Tsyrulin_JPCM_2013_DTNneutrons}%
  \BibitemOpen
  \bibfield  {author} {\bibinfo {author} {\bibfnamefont {N.}~\bibnamefont
  {Tsyrulin}}, \bibinfo {author} {\bibfnamefont {C.~D.}\ \bibnamefont
  {Batista}}, \bibinfo {author} {\bibfnamefont {V.~S.}\ \bibnamefont {Zapf}},
  \bibinfo {author} {\bibfnamefont {M.}~\bibnamefont {Jaime}}, \bibinfo
  {author} {\bibfnamefont {B.~R.}\ \bibnamefont {Hansen}}, \bibinfo {author}
  {\bibfnamefont {C.}~\bibnamefont {Niedermayer}}, \bibinfo {author}
  {\bibfnamefont {K.~C.}\ \bibnamefont {Rule}}, \bibinfo {author}
  {\bibfnamefont {K.}~\bibnamefont {Habicht}}, \bibinfo {author} {\bibfnamefont
  {K.}~\bibnamefont {Prokes}}, \bibinfo {author} {\bibfnamefont
  {K.}~\bibnamefont {Kiefer}}, \bibinfo {author} {\bibfnamefont
  {E.}~\bibnamefont {Ressouche}}, \bibinfo {author} {\bibfnamefont
  {A.}~\bibnamefont {Paduan-Filho}}, \ and\ \bibinfo {author} {\bibfnamefont
  {M.}~\bibnamefont {Kenzelmann}},\ }\bibfield  {title} {\enquote {\bibinfo
  {title} {{Neutron study of the magnetism in
  ${\mathrm{NiCl}}_{2}\mathrm{\text{-}}4\mathrm{SC}({\mathrm{NH}}_{2}{)}_{2}$}},}\
  }\href {\doibase 10.1088/0953-8984/25/21/216008} {\bibfield  {journal}
  {\bibinfo  {journal} {J. Phys.: Condens. Matter}\ }\textbf {\bibinfo {volume}
  {25}},\ \bibinfo {pages} {216008} (\bibinfo {year} {2013})}\BibitemShut
  {NoStop}%
\bibitem [{\citenamefont {Wulf}\ \emph {et~al.}(2015)\citenamefont {Wulf},
  \citenamefont {H\"{u}vonen}, \citenamefont {Sch\"{o}nemann}, \citenamefont
  {K\"{u}hne}, \citenamefont {Herrmannsd\"{o}rfer}, \citenamefont {Glavatskyy},
  \citenamefont {Gerischer}, \citenamefont {Kiefer}, \citenamefont
  {Gvasaliya},\ and\ \citenamefont
  {Zheludev}}]{WulfHuvonen_PRB_2015_DTNIntrinsicBroadening}%
  \BibitemOpen
  \bibfield  {author} {\bibinfo {author} {\bibfnamefont {E.}~\bibnamefont
  {Wulf}}, \bibinfo {author} {\bibfnamefont {D.}~\bibnamefont {H\"{u}vonen}},
  \bibinfo {author} {\bibfnamefont {R.}~\bibnamefont {Sch\"{o}nemann}},
  \bibinfo {author} {\bibfnamefont {H.}~\bibnamefont {K\"{u}hne}}, \bibinfo
  {author} {\bibfnamefont {T.}~\bibnamefont {Herrmannsd\"{o}rfer}}, \bibinfo
  {author} {\bibfnamefont {I.}~\bibnamefont {Glavatskyy}}, \bibinfo {author}
  {\bibfnamefont {S.}~\bibnamefont {Gerischer}}, \bibinfo {author}
  {\bibfnamefont {K.}~\bibnamefont {Kiefer}}, \bibinfo {author} {\bibfnamefont
  {S.}~\bibnamefont {Gvasaliya}}, \ and\ \bibinfo {author} {\bibfnamefont
  {A.}~\bibnamefont {Zheludev}},\ }\bibfield  {title} {\enquote {\bibinfo
  {title} {{Critical exponents and intrinsic broadening of the field-induced
  transition in
  ${\mathrm{NiCl}}_{2}\mathrm{\text{-}}4\mathrm{SC}({\mathrm{NH}}_{2}{)}_{2}$}},}\
  }\href {\doibase 10.1103/PhysRevB.91.014406} {\bibfield  {journal} {\bibinfo
  {journal} {Phys. Rev. B}\ }\textbf {\bibinfo {volume} {91}},\ \bibinfo
  {pages} {014406} (\bibinfo {year} {2015})}\BibitemShut {NoStop}%
\bibitem [{\citenamefont {Kenzelmann}\ \emph {et~al.}(2001)\citenamefont
  {Kenzelmann}, \citenamefont {Cowley}, \citenamefont {Buyers}, \citenamefont
  {Coldea}, \citenamefont {Gardner}, \citenamefont {Enderle}, \citenamefont
  {McMorrow},\ and\ \citenamefont {Bennington}}]{Kenzelmann_PRL_2001_CsNiCl3}%
  \BibitemOpen
  \bibfield  {author} {\bibinfo {author} {\bibfnamefont {M.}~\bibnamefont
  {Kenzelmann}}, \bibinfo {author} {\bibfnamefont {R.~A.}\ \bibnamefont
  {Cowley}}, \bibinfo {author} {\bibfnamefont {W.~J.~L.}\ \bibnamefont
  {Buyers}}, \bibinfo {author} {\bibfnamefont {R.}~\bibnamefont {Coldea}},
  \bibinfo {author} {\bibfnamefont {J.~S.}\ \bibnamefont {Gardner}}, \bibinfo
  {author} {\bibfnamefont {M.}~\bibnamefont {Enderle}}, \bibinfo {author}
  {\bibfnamefont {D.~F.}\ \bibnamefont {McMorrow}}, \ and\ \bibinfo {author}
  {\bibfnamefont {S.~M.}\ \bibnamefont {Bennington}},\ }\bibfield  {title}
  {\enquote {\bibinfo {title} {{Multiparticle States in the $\mathit{S}=1$
  Chain System ${\mathrm{CsNiCl}}_{3}$}},}\ }\href {\doibase
  10.1103/PhysRevLett.87.017201} {\bibfield  {journal} {\bibinfo  {journal}
  {Phys. Rev. Lett.}\ }\textbf {\bibinfo {volume} {87}},\ \bibinfo {pages}
  {017201} (\bibinfo {year} {2001})}\BibitemShut {NoStop}%
\bibitem [{\citenamefont {Zaliznyak}\ \emph {et~al.}(2001)\citenamefont
  {Zaliznyak}, \citenamefont {Lee},\ and\ \citenamefont
  {Petrov}}]{Zaliznyak_PRL_2001_CsNiCl3}%
  \BibitemOpen
  \bibfield  {author} {\bibinfo {author} {\bibfnamefont {I.~A.}\ \bibnamefont
  {Zaliznyak}}, \bibinfo {author} {\bibfnamefont {S.-H.}\ \bibnamefont {Lee}},
  \ and\ \bibinfo {author} {\bibfnamefont {S.~V.}\ \bibnamefont {Petrov}},\
  }\bibfield  {title} {\enquote {\bibinfo {title} {{Continuum in the
  Spin-Excitation Spectrum of a Haldane Chain Observed by Neutron Scattering in
  ${\mathrm{CsNiCl}}_{3}$}},}\ }\href {\doibase 10.1103/PhysRevLett.87.017202}
  {\bibfield  {journal} {\bibinfo  {journal} {Phys. Rev. Lett.}\ }\textbf
  {\bibinfo {volume} {87}},\ \bibinfo {pages} {017202} (\bibinfo {year}
  {2001})}\BibitemShut {NoStop}%
\bibitem [{\citenamefont {Haldane}(1983)}]{Haldane_PLA_1981_Gap}%
  \BibitemOpen
  \bibfield  {author} {\bibinfo {author} {\bibfnamefont {F.~D.~M.}\
  \bibnamefont {Haldane}},\ }\bibfield  {title} {\enquote {\bibinfo {title}
  {{Continuum dynamics of the 1-D Heisenberg antiferromagnet: identification
  with the O(3) nonlinear sigma model}},}\ }\href {\doibase
  10.1016/0375-9601(83)90631-X} {\bibfield  {journal} {\bibinfo  {journal}
  {Phys. Lett. A}\ }\textbf {\bibinfo {volume} {93}},\ \bibinfo {pages} {464}
  (\bibinfo {year} {1983})}\BibitemShut {NoStop}%
\bibitem [{\citenamefont {Sakai}\ and\ \citenamefont
  {Takahashi}(1990)}]{SakaiTakahashi_PRB_1990_DTNlikeGS}%
  \BibitemOpen
  \bibfield  {author} {\bibinfo {author} {\bibfnamefont {T.}~\bibnamefont
  {Sakai}}\ and\ \bibinfo {author} {\bibfnamefont {M.}~\bibnamefont
  {Takahashi}},\ }\bibfield  {title} {\enquote {\bibinfo {title} {{Effect of
  the Haldane gap on quasi-one-dimensional systems}},}\ }\href {\doibase
  10.1103/PhysRevB.42.4537} {\bibfield  {journal} {\bibinfo  {journal} {Phys.
  Rev. B}\ }\textbf {\bibinfo {volume} {42}},\ \bibinfo {pages} {4537}
  (\bibinfo {year} {1990})}\BibitemShut {NoStop}%
\bibitem [{\citenamefont {Yu}\ \emph {et~al.}(2012)\citenamefont {Yu},
  \citenamefont {Yin}, \citenamefont {Sullivan}, \citenamefont {Xia},
  \citenamefont {Huan}, \citenamefont {Paduan-Filho}, \citenamefont {{Oliveira
  Jr}}, \citenamefont {Haas}, \citenamefont {Steppke}, \citenamefont {Miclea},
  \citenamefont {Weickert}, \citenamefont {Movshovich}, \citenamefont {Mun},
  \citenamefont {Scott}, \citenamefont {Zapf},\ and\ \citenamefont
  {Roscilde}}]{YuYin_Nat_2012_DTNboseglass}%
  \BibitemOpen
  \bibfield  {author} {\bibinfo {author} {\bibfnamefont {R.}~\bibnamefont
  {Yu}}, \bibinfo {author} {\bibfnamefont {L.}~\bibnamefont {Yin}}, \bibinfo
  {author} {\bibfnamefont {N.~S.}\ \bibnamefont {Sullivan}}, \bibinfo {author}
  {\bibfnamefont {J.~S.}\ \bibnamefont {Xia}}, \bibinfo {author} {\bibfnamefont
  {C.}~\bibnamefont {Huan}}, \bibinfo {author} {\bibfnamefont {A.}~\bibnamefont
  {Paduan-Filho}}, \bibinfo {author} {\bibfnamefont {N.~F.}\ \bibnamefont
  {{Oliveira Jr}}}, \bibinfo {author} {\bibfnamefont {S.}~\bibnamefont {Haas}},
  \bibinfo {author} {\bibfnamefont {A.}~\bibnamefont {Steppke}}, \bibinfo
  {author} {\bibfnamefont {C.~F.}\ \bibnamefont {Miclea}}, \bibinfo {author}
  {\bibfnamefont {F.}~\bibnamefont {Weickert}}, \bibinfo {author}
  {\bibfnamefont {R.}~\bibnamefont {Movshovich}}, \bibinfo {author}
  {\bibfnamefont {E.-D.}\ \bibnamefont {Mun}}, \bibinfo {author} {\bibfnamefont
  {B.~L.}\ \bibnamefont {Scott}}, \bibinfo {author} {\bibfnamefont {V.~S.}\
  \bibnamefont {Zapf}}, \ and\ \bibinfo {author} {\bibfnamefont
  {T.}~\bibnamefont {Roscilde}},\ }\bibfield  {title} {\enquote {\bibinfo
  {title} {{Bose glass and Mott glass of quasiparticles in a doped quantum
  magnet}},}\ }\href {\doibase 10.1038/nature11406} {\bibfield  {journal}
  {\bibinfo  {journal} {Nature}\ }\textbf {\bibinfo {volume} {489}},\ \bibinfo
  {pages} {379} (\bibinfo {year} {2012})}\BibitemShut {NoStop}%
\bibitem [{\citenamefont {Wulf}\ \emph {et~al.}(2013)\citenamefont {Wulf},
  \citenamefont {H\"{u}vonen}, \citenamefont {Kim}, \citenamefont
  {Paduan-Filho}, \citenamefont {Ressouche}, \citenamefont {Gvasaliya},
  \citenamefont {Zapf},\ and\ \citenamefont
  {Zheludev}}]{WulfHuvonen_PRB_2013_DTNXdiffraction}%
  \BibitemOpen
  \bibfield  {author} {\bibinfo {author} {\bibfnamefont {E.}~\bibnamefont
  {Wulf}}, \bibinfo {author} {\bibfnamefont {D.}~\bibnamefont {H\"{u}vonen}},
  \bibinfo {author} {\bibfnamefont {J.-W.}\ \bibnamefont {Kim}}, \bibinfo
  {author} {\bibfnamefont {A.}~\bibnamefont {Paduan-Filho}}, \bibinfo {author}
  {\bibfnamefont {E.}~\bibnamefont {Ressouche}}, \bibinfo {author}
  {\bibfnamefont {S.}~\bibnamefont {Gvasaliya}}, \bibinfo {author}
  {\bibfnamefont {V.}~\bibnamefont {Zapf}}, \ and\ \bibinfo {author}
  {\bibfnamefont {A.}~\bibnamefont {Zheludev}},\ }\bibfield  {title} {\enquote
  {\bibinfo {title} {Criticality in a disordered quantum antiferromagnet
  studied by neutron diffraction},}\ }\href {\doibase
  10.1103/PhysRevB.88.174418} {\bibfield  {journal} {\bibinfo  {journal} {Phys.
  Rev. B}\ }\textbf {\bibinfo {volume} {88}},\ \bibinfo {pages} {174418}
  (\bibinfo {year} {2013})}\BibitemShut {NoStop}%
\bibitem [{\citenamefont {Lopez-Castro}\ and\ \citenamefont
  {Truter}(1963)}]{LopezCastroTruter_JChemS_1963_DTNxray}%
  \BibitemOpen
  \bibfield  {author} {\bibinfo {author} {\bibfnamefont {A.}~\bibnamefont
  {Lopez-Castro}}\ and\ \bibinfo {author} {\bibfnamefont {M.~R.}\ \bibnamefont
  {Truter}},\ }\bibfield  {title} {\enquote {\bibinfo {title} {{The crystal and
  molecular structure of dichlorotetrakisthioureanickel,
  [(NH$_2$)$_2$CS]$_4$NiCl$_2$}},}\ }\href {\doibase 10.1039/jr9630001309}
  {\bibfield  {journal} {\bibinfo  {journal} {J. Chem. Soc.}\ ,\ \bibinfo
  {pages} {1309}} (\bibinfo {year} {1963})}\BibitemShut {NoStop}%
\bibitem [{\citenamefont {Squires}(2012)}]{Squires_2012_Neutronbook}%
  \BibitemOpen
  \bibfield  {author} {\bibinfo {author} {\bibfnamefont {G.~L.}\ \bibnamefont
  {Squires}},\ }\href
  {http://www.cambridge.org/ch/academic/subjects/physics/condensed-matter-physics-nanoscience-and-mesoscopic-physics/introduction-theory-thermal-neutron-scattering-3rd-edition?format=PB}
  {\emph {\bibinfo {title} {Introduction to the Theory of Thermal Neutron
  Scattering}}}\ (\bibinfo  {publisher} {Cambridge University Press, Cambridge,
  U.K.},\ \bibinfo {year} {2012})\BibitemShut {NoStop}%
\bibitem [{\citenamefont {Prince}(2004)}]{Prince_2004_XtalTables}%
  \BibitemOpen
  \bibfield  {author} {\bibinfo {author} {\bibfnamefont {E.}~\bibnamefont
  {Prince}},\ }\href
  {http://eu.wiley.com/WileyCDA/WileyTitle/productCd-0470710292.html} {\emph
  {\bibinfo {title} {International Tables for Crystallography, Volume C:
  Mathematical, Physical and Chemical Tables}}},\ International Tables for
  Crystallography\ (\bibinfo  {publisher} {Wiley, U.K.},\ \bibinfo {year}
  {2004})\BibitemShut {NoStop}%
\bibitem [{\citenamefont {Paduan-Filho}\ \emph {et~al.}(2004)\citenamefont
  {Paduan-Filho}, \citenamefont {Gratens},\ and\ \citenamefont
  {Oliveira}}]{PaduanFilho_PRB_2004_DTNinfield}%
  \BibitemOpen
  \bibfield  {author} {\bibinfo {author} {\bibfnamefont {A.}~\bibnamefont
  {Paduan-Filho}}, \bibinfo {author} {\bibfnamefont {X.}~\bibnamefont
  {Gratens}}, \ and\ \bibinfo {author} {\bibfnamefont {N.~F.}\ \bibnamefont
  {Oliveira}},\ }\bibfield  {title} {\enquote {\bibinfo {title} {{Field-induced
  magnetic ordering in
  ${\mathrm{NiCl}}_{2}\cdot{}4\mathrm{SC}({\mathrm{NH}}_{2}{)}_{2}$}},}\ }\href
  {\doibase 10.1103/PhysRevB.69.020405} {\bibfield  {journal} {\bibinfo
  {journal} {Phys. Rev. B}\ }\textbf {\bibinfo {volume} {69}},\ \bibinfo
  {pages} {020405} (\bibinfo {year} {2004})}\BibitemShut {NoStop}%
\bibitem [{\citenamefont {Ollivier}\ and\ \citenamefont
  {Mutka}(2011)}]{OllivierMutka_JPSJ_2011_IN5}%
  \BibitemOpen
  \bibfield  {author} {\bibinfo {author} {\bibfnamefont {J.}~\bibnamefont
  {Ollivier}}\ and\ \bibinfo {author} {\bibfnamefont {H.}~\bibnamefont
  {Mutka}},\ }\bibfield  {title} {\enquote {\bibinfo {title} {{IN5 cold neutron
  time-of-flight spectrometer, prepared to tackle single crystal
  spectroscopy}},}\ }\href {\doibase 10.1143/JPSJS.80SB.SB003} {\bibfield
  {journal} {\bibinfo  {journal} {J. Phys. Soc. Jap.}\ }\textbf {\bibinfo
  {volume} {80}},\ \bibinfo {pages} {SB003} (\bibinfo {year}
  {2011})}\BibitemShut {NoStop}%
\bibitem [{Note3()}]{Note3}%
  \BibitemOpen
  \bibinfo {note} {The uniform background is slightly higher in $E_{\protect
  \text {i}}=1.17$~meV dataset, and this is the origin of the visual
  ``cut-off'' at $0.7$~meV in Fig.~\ref {FIG:TetraZone_combo}.}\BibitemShut
  {Stop}%
\bibitem [{\citenamefont {Lowde}(1960)}]{Lowde_JNE_1960_TOFresolution}%
  \BibitemOpen
  \bibfield  {author} {\bibinfo {author} {\bibfnamefont {R.~D.}\ \bibnamefont
  {Lowde}},\ }\bibfield  {title} {\enquote {\bibinfo {title} {The principles of
  mechanical neutron-velocity selection},}\ }\href {\doibase
  10.1016/0368-3265(60)90017-7} {\bibfield  {journal} {\bibinfo  {journal} {J.
  Nucl. Energy, Part A: Reactor Science}\ }\textbf {\bibinfo {volume} {11}},\
  \bibinfo {pages} {69} (\bibinfo {year} {1960})}\BibitemShut {NoStop}%
\bibitem [{\citenamefont {Lechner}(1985)}]{Lechner_1985_TOFresolution}%
  \BibitemOpen
  \bibfield  {author} {\bibinfo {author} {\bibfnamefont {R.~E.}\ \bibnamefont
  {Lechner}},\ }\bibfield  {title} {\enquote {\bibinfo {title} {{Resolution and
  intensity of a TOF-TOF spectrometer}},}\ }in\ \href
  {https://inis.iaea.org/search/search.aspx?orig_q=RN:16078322} {\emph
  {\bibinfo {booktitle} {Neutron scattering in the nineties}}}\ (\bibinfo
  {year} {1985})\BibitemShut {NoStop}%
\bibitem [{\citenamefont {Jensen}\ and\ \citenamefont
  {Mackintosh}(1991)}]{JensenMackintosh_1991_RareEarthBook}%
  \BibitemOpen
  \bibfield  {author} {\bibinfo {author} {\bibfnamefont {J.}~\bibnamefont
  {Jensen}}\ and\ \bibinfo {author} {\bibfnamefont {A.~R.}\ \bibnamefont
  {Mackintosh}},\ }\href {https://books.google.ch/books?id=LbTvAAAAMAAJ} {\emph
  {\bibinfo {title} {Rare earth magnetism: structures and excitations}}},\
  International series of monographs on physics\ (\bibinfo  {publisher}
  {Clarendon Press, UK},\ \bibinfo {year} {1991})\BibitemShut {NoStop}%
\bibitem [{\citenamefont {Zhang}\ \emph {et~al.}(2013)\citenamefont {Zhang},
  \citenamefont {Wierschem}, \citenamefont {Yap}, \citenamefont {Kato},
  \citenamefont {Batista},\ and\ \citenamefont
  {Sengupta}}]{ZhangWierschem_PRB_2013_DTNdispersion}%
  \BibitemOpen
  \bibfield  {author} {\bibinfo {author} {\bibfnamefont {Z.}~\bibnamefont
  {Zhang}}, \bibinfo {author} {\bibfnamefont {K.}~\bibnamefont {Wierschem}},
  \bibinfo {author} {\bibfnamefont {I.}~\bibnamefont {Yap}}, \bibinfo {author}
  {\bibfnamefont {Y.}~\bibnamefont {Kato}}, \bibinfo {author} {\bibfnamefont
  {C.~D.}\ \bibnamefont {Batista}}, \ and\ \bibinfo {author} {\bibfnamefont
  {P.}~\bibnamefont {Sengupta}},\ }\bibfield  {title} {\enquote {\bibinfo
  {title} {Phase diagram and magnetic excitations of anisotropic spin-one
  magnets},}\ }\href {\doibase 10.1103/PhysRevB.87.174405} {\bibfield
  {journal} {\bibinfo  {journal} {Phys. Rev. B}\ }\textbf {\bibinfo {volume}
  {87}},\ \bibinfo {pages} {174405} (\bibinfo {year} {2013})}\BibitemShut
  {NoStop}%
\bibitem [{\citenamefont {Wierschem}\ and\ \citenamefont
  {Sengupta}(2014{\natexlab{a}})}]{WierschemSengupta_PRL_2014_DTNlikeGS}%
  \BibitemOpen
  \bibfield  {author} {\bibinfo {author} {\bibfnamefont {K.}~\bibnamefont
  {Wierschem}}\ and\ \bibinfo {author} {\bibfnamefont {P.}~\bibnamefont
  {Sengupta}},\ }\bibfield  {title} {\enquote {\bibinfo {title} {{Quenching the
  Haldane Gap in Spin-1 Heisenberg Antiferromagnets}},}\ }\href {\doibase
  10.1103/PhysRevLett.112.247203} {\bibfield  {journal} {\bibinfo  {journal}
  {Phys. Rev. Lett.}\ }\textbf {\bibinfo {volume} {112}},\ \bibinfo {pages}
  {247203} (\bibinfo {year} {2014}{\natexlab{a}})}\BibitemShut {NoStop}%
\bibitem [{\citenamefont {Wierschem}\ and\ \citenamefont
  {Sengupta}(2014{\natexlab{b}})}]{WierschemSengupta_MPhysLettB_2014_DTNlikeGS}%
  \BibitemOpen
  \bibfield  {author} {\bibinfo {author} {\bibfnamefont {K.}~\bibnamefont
  {Wierschem}}\ and\ \bibinfo {author} {\bibfnamefont {P.}~\bibnamefont
  {Sengupta}},\ }\bibfield  {title} {\enquote {\bibinfo {title}
  {{Characterizing the Haldane phase in quasi-one-dimensional spin-1 Heisenberg
  antiferromagnets}},}\ }\href {\doibase 10.1142/S0217984914300178} {\bibfield
  {journal} {\bibinfo  {journal} {Mod. Phys. Lett. B}\ }\textbf {\bibinfo
  {volume} {28}},\ \bibinfo {pages} {1430017} (\bibinfo {year}
  {2014}{\natexlab{b}})}\BibitemShut {NoStop}%
\bibitem [{\citenamefont {Pintschovius}\ \emph {et~al.}(2014)\citenamefont
  {Pintschovius}, \citenamefont {Reznik}, \citenamefont {Weber}, \citenamefont
  {Bourges}, \citenamefont {Parshall}, \citenamefont {Mittal}, \citenamefont
  {Chaplot}, \citenamefont {Heid}, \citenamefont {Wolf}, \citenamefont
  {Lamago},\ and\ \citenamefont {Lynn}}]{Pintschovius_JAC_2014_18mevSpurion}%
  \BibitemOpen
  \bibfield  {author} {\bibinfo {author} {\bibfnamefont {L.}~\bibnamefont
  {Pintschovius}}, \bibinfo {author} {\bibfnamefont {D.}~\bibnamefont
  {Reznik}}, \bibinfo {author} {\bibfnamefont {F.}~\bibnamefont {Weber}},
  \bibinfo {author} {\bibfnamefont {P.}~\bibnamefont {Bourges}}, \bibinfo
  {author} {\bibfnamefont {D.}~\bibnamefont {Parshall}}, \bibinfo {author}
  {\bibfnamefont {R.}~\bibnamefont {Mittal}}, \bibinfo {author} {\bibfnamefont
  {S.~L.}\ \bibnamefont {Chaplot}}, \bibinfo {author} {\bibfnamefont
  {R.}~\bibnamefont {Heid}}, \bibinfo {author} {\bibfnamefont {T.}~\bibnamefont
  {Wolf}}, \bibinfo {author} {\bibfnamefont {D.}~\bibnamefont {Lamago}}, \ and\
  \bibinfo {author} {\bibfnamefont {J.~W.}\ \bibnamefont {Lynn}},\ }\bibfield
  {title} {\enquote {\bibinfo {title} {{Spurious peaks arising from multiple
  scattering events involving the sample environment in inelastic neutron
  scattering}},}\ }\href {\doibase 10.1107/S1600576714010140} {\bibfield
  {journal} {\bibinfo  {journal} {J. Appl. Cryst.}\ }\textbf {\bibinfo {volume}
  {47}},\ \bibinfo {pages} {1472} (\bibinfo {year} {2014})}\BibitemShut
  {NoStop}%
\bibitem [{\citenamefont {Utesov}\ \emph {et~al.}(2014)\citenamefont {Utesov},
  \citenamefont {Sizanov},\ and\ \citenamefont
  {Syromyatnikov}}]{UtesovSizanov_PRB_2014_DisorderExcitations}%
  \BibitemOpen
  \bibfield  {author} {\bibinfo {author} {\bibfnamefont {O.~I.}\ \bibnamefont
  {Utesov}}, \bibinfo {author} {\bibfnamefont {A.~V.}\ \bibnamefont {Sizanov}},
  \ and\ \bibinfo {author} {\bibfnamefont {A.~V.}\ \bibnamefont
  {Syromyatnikov}},\ }\bibfield  {title} {\enquote {\bibinfo {title} {Localized
  and propagating excitations in gapped phases of spin systems with bond
  disorder},}\ }\href {\doibase 10.1103/PhysRevB.90.155121} {\bibfield
  {journal} {\bibinfo  {journal} {Phys. Rev. B}\ }\textbf {\bibinfo {volume}
  {90}},\ \bibinfo {pages} {155121} (\bibinfo {year} {2014})}\BibitemShut
  {NoStop}%
\bibitem [{Note4()}]{Note4}%
  \BibitemOpen
  \bibinfo {note} {{D. H\"{u}vonen \protect \emph {et al.},
  unpublished}}\BibitemShut {NoStop}%
\bibitem [{Note5()}]{Note5}%
  \BibitemOpen
  \bibinfo {note} {Ref.~\protect \rev@citealpnum {Vojta_PRL_2013_InGap},
  Fig.~1}\BibitemShut {NoStop}%
\bibitem [{Note6()}]{Note6}%
  \BibitemOpen
  \bibinfo {note} {Ref.~\protect \rev@citealpnum {Vojta_PRL_2013_InGap}
  Supplementary Material, Fig.~S2}\BibitemShut {NoStop}%
\bibitem [{\citenamefont {Indenbom}\ and\ \citenamefont
  {Lothe}(1992)}]{IndenbomLothe_1992_DefectBook}%
  \BibitemOpen
  \bibfield  {author} {\bibinfo {author} {\bibfnamefont {V.~L.}\ \bibnamefont
  {Indenbom}}\ and\ \bibinfo {author} {\bibfnamefont {J.}~\bibnamefont
  {Lothe}},\ }\href
  {http://store.elsevier.com/Elastic-Strain-Fields-and-Dislocation-Mobility/isbn-9780444600424/}
  {\emph {\bibinfo {title} {Elastic Strain Fields and Dislocation Mobility}}},\
  Modern Problems in Condensed Matter Sciences\ (\bibinfo  {publisher} {North
  Holland, Netherlands},\ \bibinfo {year} {1992})\BibitemShut {NoStop}%
\bibitem [{\citenamefont {Zvyagin}\ \emph {et~al.}(2008)\citenamefont
  {Zvyagin}, \citenamefont {Wosnitza}, \citenamefont {Kolezhuk}, \citenamefont
  {Zapf}, \citenamefont {Jaime}, \citenamefont {Paduan-Filho}, \citenamefont
  {Glazkov}, \citenamefont {Sosin},\ and\ \citenamefont
  {Smirnov}}]{Zvyagin_PRB_208_ESRinDTN}%
  \BibitemOpen
  \bibfield  {author} {\bibinfo {author} {\bibfnamefont {S.~A.}\ \bibnamefont
  {Zvyagin}}, \bibinfo {author} {\bibfnamefont {J.}~\bibnamefont {Wosnitza}},
  \bibinfo {author} {\bibfnamefont {A.~K.}\ \bibnamefont {Kolezhuk}}, \bibinfo
  {author} {\bibfnamefont {V.~S.}\ \bibnamefont {Zapf}}, \bibinfo {author}
  {\bibfnamefont {M.}~\bibnamefont {Jaime}}, \bibinfo {author} {\bibfnamefont
  {A.}~\bibnamefont {Paduan-Filho}}, \bibinfo {author} {\bibfnamefont {V.~N.}\
  \bibnamefont {Glazkov}}, \bibinfo {author} {\bibfnamefont {S.~S.}\
  \bibnamefont {Sosin}}, \ and\ \bibinfo {author} {\bibfnamefont {A.~I.}\
  \bibnamefont {Smirnov}},\ }\bibfield  {title} {\enquote {\bibinfo {title}
  {{Spin dynamics of
  $\mathrm{Ni}{\mathrm{Cl}}_{2}\text{-}4\mathrm{S}\mathrm{C}{(\mathrm{N}{\mathrm{H}}_{2})}_{2}$
  in the field-induced ordered phase}},}\ }\href {\doibase
  10.1103/PhysRevB.77.092413} {\bibfield  {journal} {\bibinfo  {journal} {Phys.
  Rev. B}\ }\textbf {\bibinfo {volume} {77}},\ \bibinfo {pages} {092413}
  (\bibinfo {year} {2008})}\BibitemShut {NoStop}%
\bibitem [{\citenamefont {Sizanov}\ and\ \citenamefont
  {Syromyatnikov}(2011)}]{SizanovSyromyatnikov_PRB_2011_bosons4DTN}%
  \BibitemOpen
  \bibfield  {author} {\bibinfo {author} {\bibfnamefont {A.~V.}\ \bibnamefont
  {Sizanov}}\ and\ \bibinfo {author} {\bibfnamefont {A.~V.}\ \bibnamefont
  {Syromyatnikov}},\ }\bibfield  {title} {\enquote {\bibinfo {title} {Bosonic
  representation of quantum magnets with large single-ion easy-plane
  anisotropy},}\ }\href {\doibase 10.1103/PhysRevB.84.054445} {\bibfield
  {journal} {\bibinfo  {journal} {Phys. Rev. B}\ }\textbf {\bibinfo {volume}
  {84}},\ \bibinfo {pages} {054445} (\bibinfo {year} {2011})}\BibitemShut
  {NoStop}%
\end{thebibliography}%
\end{document}